\documentclass[aps, prx, amsmath, amssymb, a4paper, floatfix, superscriptaddress, longbibliography]{revtex4-2}

\usepackage{algorithm}
\usepackage{algorithmicx}
\usepackage{algpseudocode}
\usepackage{amsfonts}
\usepackage{amsthm}
\usepackage[title]{appendix}
\usepackage[english]{babel}
\usepackage{bm}
\usepackage{booktabs}
\usepackage{comment}
\usepackage{csquotes}
\usepackage{graphicx} 
\usepackage[utf8x]{inputenc} 
\usepackage{listings}
\usepackage{mathrsfs}
\usepackage[version=4]{mhchem}
\usepackage{multirow}
\usepackage{physics}
\usepackage{siunitx}
\usepackage{textcomp}
\usepackage{textgreek}
\usepackage[normalem]{ulem}
\usepackage{xcolor}

\begin{document}

\title[]{Cavitation by phase shift of focused shock waves inside a droplet}




\author{Samuele Fiorini} 
    \email{sfiorini@ethz.ch}

\affiliation{Institute of Fluid Dynamics, D-MAVT, ETH Zürich, Sonneggstrasse 3, Zürich, 8092, Switzerland}

\author{Guillaume T. Bokman}

\altaffiliation{These authors contributed equally to this work.}

\affiliation{Institute of Fluid Dynamics, D-MAVT, ETH Zürich, Sonneggstrasse 3, Zürich, 8092, Switzerland}

\affiliation{Plasma Physics Group, Imperial College London, London SW7 2BW, United Kingdom}

\author{Anunay Prasanna}
\altaffiliation{These authors contributed equally to this work.}

\affiliation{Institute of Fluid Dynamics, D-MAVT, ETH Zürich, Sonneggstrasse 3, Zürich, 8092, Switzerland}

\author{Stefanos Nikolaou}

\affiliation{Institute of Fluid Dynamics, D-MAVT, ETH Zürich, Sonneggstrasse 3, Zürich, 8092, Switzerland}

\author{Sayaka Ichihara}
\affiliation{Department of Mechanical Systems Engineering, Tokyo University of Agriculture and Technology, Nakacho 2-24-16, Koganei, 184-8588, Tokyo, Japan}
\affiliation{Faculty of Natural Sciences, Institute for Physics, Otto-von-Guericke-University Magdeburg, Universitätsplatz 2, 39106, Magdeburg, Germany}

\author{Bratislav Lukić}

\affiliation{ESRF -- The European Synchrotron, CS 40220, Grenoble, F-38043, France}

\affiliation{Institut Max von Laue - Paul Langevin (ILL) Avenue des Martyrs 71, Grenoble, 38000, France}

\author{Alexander Rack}
\affiliation{ESRF -- The European Synchrotron, CS 40220, Grenoble, F-38043, France}

\author{Yoshiyuki Tagawa}

\affiliation{Department of Mechanical Systems Engineering, Tokyo University of Agriculture and Technology, Nakacho 2-24-16, Koganei, 184-8588, Tokyo, Japan}

\author{Outi Supponen}
    \email{outis@ethz.ch}

\affiliation{Institute of Fluid Dynamics, D-MAVT, ETH Zürich, Sonneggstrasse 3, Zürich, 8092, Switzerland}

\begin{abstract}
Localized cavitation in liquids and soft tissues, typically initiated by the rarefaction phase of high-amplitude ultrasound waves, is leveraged in several biomedical applications such as ablation techniques and drug delivery with vaporizing agents. 
However, safety considerations aimed at avoiding unwanted bubble activity outside the targeted region pose a limit to the maximum allowed peak negative pressure, which on the other hand can hinder the therapeutic efficacy of these techniques. 
This study shows that a purely compressive shock wave can generate localized negative pressure and initiate cavitation inside a sub‑millimetric perfluorohexane droplet, without requiring any externally applied rarefaction wave.
The Gouy phase shift is identified as the physical mechanism responsible for the conversion of positive pressure into tension during shock focusing, and its occurrence is demonstrated through numerical simulations and direct experimental measurements. 
Comparison of the regions affected by cavitation, visualized \emph{in-situ} by means of high-speed x-ray phase-contrast imaging, with prediction from Classical Nucleation Theory suggests homogeneous nucleation as the underlying mechanism behind bubble formation.
The presented findings offer valuable insights into the physics of shock wave propagation which can inspire the development of novel acoustic driving strategies for cavitation generation, facilitating the reduction of negative pressures outside the target region and improving the safety and precision of biomedical treatments. 

\end{abstract}


\maketitle

\section{Introduction} \label{sec:Intro}
Cavitation is a physical phenomenon comprising the nucleation, growth, oscillation, and collapse of bubbles in a liquid.
Traditionally, the term has been used to describe the formation of vapor bubbles in high-speed water flows, such as in hydraulic machinery like centrifugal pumps, water turbines and boat propellers.
The ability to generate controlled cavitation using acoustic and shock waves has opened up the possibility of exploiting this phenomenon for cleaning~\cite{fuchs_ultrasonic_2015}, enhancement of chemical reactions~\cite{kiss_ultrasoundassisted_2018}, and non-invasive medical treatments such as extracorporeal shock wave therapy (ESWT)~\cite{simplicio_extracorporeal_2020}, extracorporeal shock wave lithotripsy (ESWL)~\cite{qiang_effectiveness_2020}, histotrispy~\cite{xu_histotripsy_2021} and Blood-Brain Barrier (BBB) opening \cite{liao_investigation_2021}. 
Moreover, acoustically-induced cavitation in superheated micro- and nanodroplets, commonly referred to as acoustic droplet vaporization (ADV), is being studied due to its potential applications in techniques such as embolotherapy~\cite{harmon_minimally_2019} and targeted drug delivery~\cite{chen_targeted_2013, xiao_acoustically_2024, abeid_real-time_2025}.

Cavitation inception can occur when the liquid is brought below its saturation pressure, which is a necessary condition for the formation and growth of vapor bubbles to become thermodynamically favorable~\cite{carey_liquid-vapor_2020}. 
Since the presence of intermolecular bonds allows liquids to counteract pulling forces~\cite{rzoska_how_2007}, they can sustain pressures below $ 0\ \si{\pascal} $ without undergoing phase change, in a condition of so-called metastability.  
If the negative pressure, commonly called tension, is strong enough, the thermodynamic stability limit of the system is reached and the phase transition happens abruptly. Classical Nucleation Theory (CNT), combined with the hypothesis of homogeneous nucleation, provides a theoretical framework that can model the formation and the stability of a pure vapor bubble in terms of Gibbs Free Energy variation~\cite{carey_liquid-vapor_2020}.
However, in most practical applications the formation of bubbles happens before the stability limit of the system is reached. The presence of tiny, stabilized gas pockets in the liquid that can grow and eventually collapse upon application of a rarefaction pressure, in a process called heterogeneous cavitation~\cite{mancia_acoustic_2021}, is a commonly accepted explanation for this behavior.


Ultrasound transducers are one of the simplest ways to generate the low pressures required for controlled cavitation. Typically, these devices consist of one or more piezoelectric elements whose mechanical vibrations create an acoustic wave in the surrounding fluid \cite{cobbold_foundations_2023}. If the source is properly shaped and the excitation frequency is high enough, the wave can be focused at a certain distance from the transducer with the goal of constraining the high amplitude waves in a narrow region, a feature particularly important in biomedical applications where off-target cavitation activity can damage healthy tissue. Indeed, safety considerations impose some of the main restrictions on the maximum admissible pressure amplitude for therapeutic use. For example, the peak rarefaction pressure of the ultrasound wave, divided by the square root of its center frequency, should be limited to avoid transient cavitation of gas pockets dissolved in tissue and blood flow~\cite{apfel_gauging_1991} in applications such as ultrasound-targeted drug delivery with contrast agents or ultrasound imaging.
Even in treatments where cavitation activity generates the most therapeutic effect and an increase in negative pressure at the target location can be beneficial, the maximum rarefaction pressure must be limited to reduce collateral damage outside the focal region.
The possibility of generating the required negative pressure directly at the target location would therefore offer a way to improve the efficacy of the treatments while complying with the safety limitations.  


As an alternative, the reflection of positive pressure waves against an acoustically soft interface can be exploited to generate localized tension without the need of negative pressure waves traveling in the surrounding fluid. If an acoustic wave is reflected at an interface when traveling from a material with a higher acoustic impedance to one with a lower acoustic impedance, the resulting reflection coefficient will be negative \cite{kinsler_fundamentals_2000}. Therefore, the reflected wave will have the opposite sign with respect to the incident wave (i.e. it will acquire a phase shift of $ \pi\ \si{\radian} $). Cavitation due to reflection of positive pressure pulses, especially shock waves, has been reported by several authors both for flat~\cite{bogach_strength_2000,  boteler_tensile_2004} and curved interfaces~\cite{obreschkow_confined_2011, stan_negative_2016, huneault_shock_2019}, and is typically referred to as spallation. 

Another interesting possibility for localized negative pressure generation relies on taking advantage of the Gouy phase shift effect in 3-D focusing acoustic waves. This phenomenon, well known in the field of optics \cite{kaltenecker_gouy_2016} and recently reported for acoustic waves in liquids \cite{lee_origin_2020}, consists of a converging sinusoidal wave gradually acquiring a phase of $ \pi\ \si{\radian} $ as it crosses the focal point. 
For an arbitrarily shaped acoustic wave, such as a distorted ultrasound burst or a single pressure pulse, the phase shift of each harmonic component would result in a sign reversal of the wave shape, meaning that positive pressure can be locally converted into negative pressure.
This phenomenon has been hypothesized to contribute substantially to generate negative pressure, and even initiate cavitation, inside perfluorocarbon microdroplets interacting with ultrasound waves presenting a pronounced compression phase~\cite{fiorini_Positive_2024}. However, a clear observation of a phase shift of ultrasound waves in a confinement is challenged by the presence of multiple compression and rarefaction peaks simultaneously undergoing phase shift and by the interference between the transmitting and reflecting wavefronts. 
The use of a shock wave can provide a practical alternative to characterize the focusing process of a positive pressure pulse and to evaluate its capability to initiate cavitation. The presence of a predominant positive pressure peak in the waveform ensures that the eventual occurrence of Gouy phase shift clearly manifest in terms of negative pressure generation. 
However, a formal expression for Gouy phase shift is derived by solving the paraxial approximation of the wave equation for a Gaussian beam~\cite{gillen_light_2017}, and is therefore rigorously valid only for linear acoustics. 
Nevertheless, Gouy phase shift might also occur during shock wave focusing, since shock wave dynamics is described by the Euler equations from which the wave equation can in turn be derived~\cite{kinsler_fundamentals_2000}. 
As a matter of fact, cavitation activity has been detected in the center of a circular 2-D focusing shock and its cause has been attributed to Gouy phase shift \cite{pezeril_direct_2011, veysset_single-bubble_2018}. Furthermore, recent numerical work on shock wave focusing in the human eye during laser eye surgery has also hypothesized the occurrence phase shift of the focusing wave \cite{pozar_simulation_2018}.

In this work, we show that a purely compressive focusing shock wave in a liquid can generate strong tension around the focal region and initiate cavitation thanks to Gouy Phase shift. A sub-millimetric perfluorohexane (\ce{C6F14}, PFH) droplet immersed in water is chosen to act as an acoustic lens and to focus the shock wave. A weak shock wave (Ma $ < $ 1.1) is created by optical breakdown of water to ensure a negligible tensile tail after the main positive pressure pulse and minimize the interference between wavefronts after reflections at the droplet internal interface. Advanced experimental techniques such as high-speed x-ray phase-contrast imaging are employed to visualize cavitation activity inside the PFH droplet. Projected-background oriented schlieren (BOS) measurements are combined with numerical simulations to confirm localized tension generation within the droplet. The numerical pressure field is additionally used to model bubble formation under the assumptions of both heterogeneous cavitation and homogeneous nucleation to identify the cavitation onset mechanism. 




\section{Methods}
\subsection{Propagation-based phase-contrast x-ray imaging}
\label{sec:xrayMethods}

\begin{figure}[h]
    \centering
    \includegraphics[width = \linewidth]{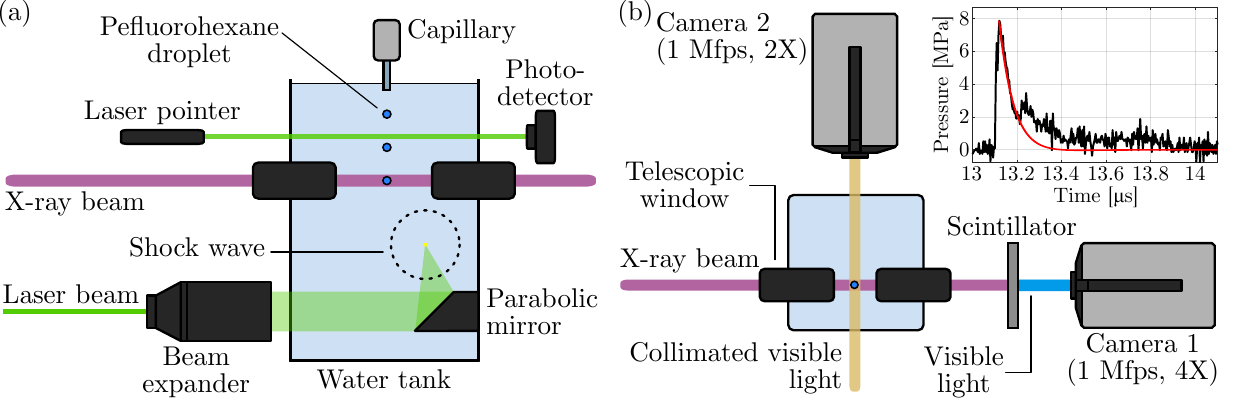}
    \caption{Combined high-speed optical and phase-contrast x-ray imaging setup. The side view (a) shows details of the laser path for the shock wave generation and positioning of the droplet-producing capillary. An overview of the light paths can be seen in the top view (b). The inset shows a typical shock wave profile measured using a 75-\si{\micro \meter} needle hydrophone located at a distance of approximately 20 \si{\milli\meter} from the shock wave source.}
    \label{fig:experimentalSetupXRay}
\end{figure}

The experimental setup employed to visualize cavitation inside the PFH droplet is illustrated in Fig.~\ref{fig:experimentalSetupXRay}. Droplets with radii ranging between 400 and 900 \si{\micro\meter} are generated with a vertically held glass capillary with an inner diameter of 0.5 \si{\milli\meter)} connected to a syringe pump (Syringe pump 33, Harvard Apparatus) and submerged in a tank filled with distilled water. The droplets are composed of PFH (APF-60HP, FluoroMed) that has a boiling temperature of 56~\si{\celsius}. The liquid has been selected among fluorocarbons because of their relevance for ADV, where this class of materials is commonly used as dispersed phase to produce droplets emulsions. Moreover, its higher boiling point with respect to lighter perfluorocarbons (29~\si{\celsius} for \ce{C5F12} for example) provides good stability against spontaneous vaporization during experiments. The shock wave is generated by optical breakdown of water using a frequency-doubled pulsed infrared Q-switched Nd:YAG laser (Quantel Q-smart 532~\si{\nano\meter}, Lumibird)~\cite{sankin_focusing_2008}. The laser beam is first expanded using a 10$\times$ beam expander (52-71-10X-532/1064, Special Optics) and then focused onto a narrow spot by an aluminum 90\textdegree\ parabolic mirror (35-508, Edmund Optics) located on the distal wall of the water tank. The tight focusing of the laser causes optical breakdown of the water at the mirror's focal point, which generates a highly pressurized plasma, as well as a spherically-propagating shock wave. 
The tank is equipped with two movable telescopic windows that can be adjusted to shorten the path of the x-rays through water in order to reduce absorption. Additional details on the tank design and the optical components can be found in \citet{bokman_high-speed_2023}. 

To overcome optical distortion of visible light due to refraction at the water-PFH interface, which would prevent visual access to the droplet interior and clear imaging of cavitation activity, propagation-based phase-contrast x-ray imaging experiments at the 150-\si{\meter} long ID19 beamline of the European Synchrotron Radiation Facility (ESRF, France) are performed. The imaging system is positioned at a distance of 7.5~\si{\meter} downstream from the sample in order to ensure that the interaction between the undisturbed and the refracted x-ray beam occurs in the near-field Fresnel diffraction regime, before the detection takes place. The ESRF's 16-bunch filling mode provides x-ray pulses of 60~\si{\pico\second} duration every 176~\si{\nano\second}. X-rays transmitted through the samples are partially absorbed by a LYSO:Ce single-crystal scintillator and re-emitted in visible light with a peak wavelength of 413~\si{\nano\meter} and approximately 40~\si{\nano\second} afterglow. The visible light is captured by an ultra-high-speed camera (HPV-X2, Shimadzu) equipped with a 4$\times$ magnification objective (0.2 numerical aperture), resulting in a sampling resolution $ d_{xy} = $ 8~\si{\micro\meter}/pixel for the x-ray videos~\cite{olbinado_mhz_2017}. The recording speed is set to 0.94~\si{\mega\hertz}. The obtained radiographs present a strong edge contrast thanks to the (partial) spatial coherent illumination~\cite{wilkins_Phase-contrast_1996, cloetens_Phase_1996}, while offering full visual access to the inside of the droplet bulk to detect nucleation events. On the other hand, the lower sensitivity of x-rays to density (and therefore refractive index) gradients with respect to visible light prevents the visualization of shock waves in the present configuration. 

The shock wave's propagation in water is imaged using a collimated visible light source (OSL2, Thorlabs) and a second ultra-high-speed camera (HPV-X2, Shimadzu) synchronized with the first one. A macro lens (2X Ultra Macro APO, Laowa) with a 2$\times$ magnification is mounted on the camera, which provides a resolution of 16~\si{\micro\meter}/pixel for the shadowgraph videos. A laser barrier, composed of a horizontal laser diode (PL251, Thorlabs) pointing at a photodetector (DET10A2, Thorlabs), is used to detect the passage of the falling droplet. The voltage reduction caused by the droplet interrupting the laser beam is used to trigger the synchronized image acquisition of the two cameras and the Nd:YAG laser. A full description of the triggering sequence can be found in Appendix~\ref{app:trigger}.

A thorough characterization of the shock wave pressure profile is performed by placing a 75-\si{\micro \meter} hydrophone (NH0075 Precision Acoustics) in the system depicted in Fig.~\ref{fig:experimentalSetupXRay} before the introduction of the capillary. The hydrophone is located approximately 20~\si{\milli\meter} from the origin of the shock wave to avoid signal saturation and damage to the sensor. 

The pressure profiles recorded at the hydrophone location are modeled using a Friedlander function \cite{sochet_friedlander_2018}:
    \begin{equation}
        p(t) = p_{\mathrm{max}, 2}(1 - t/t_I)\exp(-\beta t/t_I),
        \label{eq:friedlanderEquation}
    \end{equation}
where $ p_{\mathrm{max}, 2} $ is the measured peak pressure, $ t_I $  is the duration of the compressive part of the shock wave and $ \beta $ is a parameter that describes its exponential decay. The values for $ t_I $ and $ \beta $ are selected by Least-Square fitting.
An example of the recorded pressure signal and the corresponding Friedlander profile are shown in the inset of Fig.~\ref{fig:experimentalSetupXRay}(b), with $ t = 0 $ being the time at which the laser is fired.
It is worth noting that the laser-induced shock waves are characterized by a practically absent tensile tail, making the shock wave a good approximation of a purely compressive traveling wave. 
    
The maximum pressure $  p_{\mathrm{max}, 1} $ at the droplet location is obtained using the scaling presented in~\citet{bokman_scaling_2023}, which models the reduction of the shock wave peak positive pressure due to spherical propagation and nonlinear dissipation as a function of time:
\begin{equation}
    p_{\mathrm{max}, 1} = p_{\mathrm{max},2}(t_{0, 1}/t_{0, 2})^{-1.4},
    \label{eq:scalingLawPMax}
\end{equation}
where $ p_{\mathrm{max}, 1}$ and $ p_{\mathrm{max}, 2} $ are the pressure peaks, and $ t_{0, 1} $, $t_{0, 2} $ are the times at which the shock reaches the droplet and hydrophone location, respectively. The impact time $ t_{0, 1} = 8.26\ \pm 0.85\ \si{\micro\second} $ between the shock wave and the droplet is estimated from the shadowgraphs (a selection of whose frames is available as Supplementary Material, see Appendix~\ref{sec:suppMat}), knowing that the first recorded frame corresponds to the moment of shock wave inception. The time $ t_{0, 2} = 13.114\ \pm 0.008\ \si{\micro\second} $ is directly available from the hydrophone measurements.
The average value of the shock wave's peak positive pressure measured by the needle hydrophone is $  p_{\mathrm{max}, 2} = 7.53 \pm 0.34\ \si{\mega\pascal} $, which yields a peak positive pressure at the droplet location of $ p_{\mathrm{max}, 1} = 14.39 \pm 2.07\ \si{\mega\pascal} $ when substituted in Eq.~(\ref{eq:scalingLawPMax}) along with the values of $ t_{0, 1} $ and $ t_{0, 2} $.

The final expression for the pressure profile at the droplet location then reads:
\begin{equation}
    p(t) = p_{\mathrm{max}, 1}(1 - t/\overline{t}_I)\exp(-\overline{\beta} t/\overline{t}_I).
    \label{eq:averageFriedlanderEquation}
\end{equation}
The value of $\overline{\beta} = 3.66 \pm 0.21 $ and $ \overline{t}_I = 273 \pm 13\ \si{\nano\second} $ are calculated by averaging the values of $ \beta $ and $ t_I $ obtained from four different measurements.

\subsection{Numerical simulations}
\label{sec:simulationSetup}
In this work, numerical hydrodynamic simulations are used to visualize the formation of negative pressures inside the PFH droplet due to the focusing of the shock wave.
Due to the presence of two different liquids and the expected occurrence of high pressure (or tension) regions where the hypothesis of small perturbations is not valid, the multi-phase, diffuse-interface-method-based Euler equation solver implemented in ECOGEN \cite{schmidmayer_ecogen_2020} has been selected. This code has been validated for various multiphase compressible problems \cite{pishchalnikov_high-speed_2019, trummler_near-surface_2020, dorschner_formation_2020}. The model is described in~\citet{kapila_two-phase_2001} and \citet{richard_saurel_simple_2009} and reads:
\begin{equation}
    \begin{cases}
        \frac{\partial \alpha_j}{\partial t} + \bm{u}\cdot{\nabla \alpha_j} &= \delta p_j\\
        \frac{\partial \alpha_j \rho_j}{\partial t} + \nabla \cdot (\alpha_j \rho_j \bm{u}) &= 0\\
        \frac{\partial \rho \bm{u}}{\partial t} + \nabla \cdot (\rho \bm{u} \otimes \bm{u} + p\bm{I}) &= 0\\
        \frac{\partial \rho E}{\partial t} + \nabla \cdot [(\rho E + p) \bm{u}] &= 0\\
    \end{cases}
    \label{eq:ECOGEN}
\end{equation}
where $ j = 1, 2, \cdots N $ specify the phase, $ \alpha_j $, $ \rho_j $ and $ \delta p_j $ represent the volume fraction, density and pressure-relaxation term of the $j_{\mathrm{th}}$ phase and $ \otimes $ is the tensor product operator. The pressure and velocity vector of the mixture correspond to $ p $ and $ \bm{u} $, respectively. The mixture density and internal energy are defined as $ \rho = \sum_j\alpha_j\rho_j $ and $ e = \sum_j \alpha_j \rho_j e_j / \rho $, with $ e_j $ being the internal energy of the $ j_{\mathrm{th}} $ phase. 
The mixture total energy reads $ E = e + \frac{1}{2}||\bm{u}||^2$.

Water is modeled using the Stiffened-Gas Equation of State (SG EoS) as defined in~\citet{le_metayer_elaboration_2004}:
    \begin{align}
        e_\mathrm{w}(p_\mathrm{w}, v_\mathrm{w}) &= \left(\frac{p_\mathrm{w} + \gamma p_\mathrm{\infty, w}}{\gamma_\mathrm{w} - 1} \right)v_\mathrm{w} + e_\mathrm{ref, w} \\
        v_\mathrm{w}(p_\mathrm{w}, T_\mathrm{w}) &= \frac{(\gamma_\mathrm{w} - 1) c_\mathrm{v, w} T_\mathrm{w}}{p_\mathrm{w} + p_\mathrm{\infty, w}},
        \label{Water_EoS}
    \end{align}
where $ e_\mathrm{w} $, $ p_\mathrm{w} $, $ v_\mathrm{w} $, and $ T_\mathrm{w} $ are the internal energy, pressure, specific volume, and temperature of water. The specific heat capacity at constant volume, the adiabatic index and the internal energy reference are defined as $ c_\mathrm{v, w} $, $ \gamma_\mathrm{w} $ and $ e_\mathrm{{ref}, w}$, respectively. The stiffness parameter $ p_\mathrm{\infty, w} $ represents the ability of the material to withstand negative pressure without undergoing phase change~\cite{prasanna_shock_2025}. All the quantities $ c_\mathrm{v, w} $, $ \gamma_\mathrm{w} $, $ e_\mathrm{{ref}, w} $ and $ p_\mathrm{\infty, w} $ are assumed constant. For PFH, the Nobel-Abel-Stiffened-Gas (NASG) EoS is employed~\cite{le_metayer_noble-abel_2016}:
\begin{align}
        e_\mathrm{P}(p_\mathrm{P}, v_\mathrm{P}) = \left(\frac{p_\mathrm{P} + \gamma p_\mathrm{\infty, P}}{\gamma_\mathrm{P} - 1} \right)(v_\mathrm{P} - b_\mathrm{P}) + e_\mathrm{{ref}, P} \\
        v_\mathrm{P}(p_\mathrm{P}, T_\mathrm{P}) = \frac{(\gamma_\mathrm{P} - 1) c_{v, P} T_\mathrm{P}}{p_\mathrm{P} + p_\mathrm{\infty, P}} + b_\mathrm{P},
        \label{eq:PFH_EoS}
\end{align}
where all the quantities have the same definition as above, with the subscript `$ \mathrm{P} $' denoting PFH. Additionally, the parameter $ b_\mathrm{P} $ accounts for the repulsion between the material's molecules. The quantities $ c_\mathrm{v, P} $, $ \gamma_\mathrm{P} $, $ e_\mathrm{{ref}, P} $, $ p_\mathrm{\infty, P} $ and $ b_\mathrm{P} $ are assumed constant. 
The values of the parameters used for the simulation are summarized in Table~\ref{tab:EoS_Parameters}. The values for water are implemented in the ECOGEN database, while the ones for PFH are taken from the work of~\citet{prasanna_shock_2025}. 

\begin{table}[h]
    \caption{Thermodynamic parameters defining the equations of state for water and PFH.}
    \label{tab:EoS_Parameters}
    \begin{tabular}{l|cc}
        \ & \textrm{Water} & \textrm{PFH} \\
        \hline
        $\gamma$ [-] & 2.96 & 1.48 \\ 
        $b$ [\si{\meter\cubed\per\kilo\gram}] & - & $ 2.04\times 10^{-4}$\\ 
        $c_v$ [\si{\joule\per\kilo\gram\per\kelvin}] & 1231.2  & $ 515.6 $ \\ 
        $p_{\infty}$ [\si{\pascal}] & $ 7.22 \times 10^8 $  & $ 1.76 \times 10^8 $ \\ 
        $ e_{\mathrm{ref}} [\si{\joule\per\kilo\gram}] $ & 0 & $ -227 \times 10^3 $  \\ 
        \hline
    \end{tabular}
\end{table}

\begin{figure}[h]
    \centering
    \includegraphics[width=0.5\linewidth]{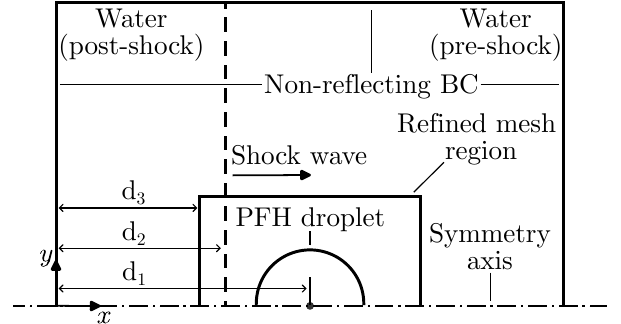}
    \caption{Schematic of the numerical domain with the boundary conditions. A spherical PFH droplet is placed on the symmetry axis at the bottom edge with the center at a distance $ d_1 $ from the left side. The post-shock state is defined in Eq.~\eqref{eq:initial_Pressure} such that the wave front is at a distance $ d_2 $ from the left edge, traveling from left to right. The start of the refined mesh region is placed at a distance $ d_3 $ from the left side of the domain. Non-reflecting boundary conditions are applied on the left, right and top edge. The origin is placed at the bottom-left corner of the domain.}
\label{fig:simulationSetup}
\end{figure}

The numerical setup shown in Fig.~\ref{fig:simulationSetup} is designed to emulate the experiments. 
For simplicity, the droplet is modeled as a sphere to enable 2D axisymmetric simulations, and the shock wave is modeled as a planar blast wave because its radius of curvature is more than an order of magnitude larger than that of the droplet at the moment of impact. 
The numerical domain consists of a 14~\si{\milli\meter}~$\times$ 7~\si{\milli\meter} rectangle of water, in which a spherical PFH droplet of radius $ R $ is placed with its centerline aligned with the bottom edge and its center at a distance $ d_1 = 7.005~\si{\milli\meter} + R $ from the left side of the domain. Non-reflective boundary conditions are applied to the left, right, and upper edges, while the lower edge is the axis of symmetry.  The origin is placed at the bottom-left corner. A cell resolution of $ \Delta x = \Delta y = 5\ \si{\micro\meter} $ is applied to a $ 4R\ \times\ 2R $ box enclosing the droplet and located at a horizontal distance $ d_3 = d_1 - 2R = 7.005~\si{\milli\meter} - R $ from origin. Mesh stretching with a geometric progression of 1.05 is applied outside of that box. 
Adaptive mesh refinement is also employed, with a maximum of 4 refinement levels \cite{schmidmayer_adaptive_2019}.

The shock wave front is initialized in water at a horizontal distance of $ d_2 = 7~\si{\milli\meter} $ from the origin. The initial location of the shock is chosen for its proximity to the PFH droplet, while ensuring that the dynamics of interest within the droplet are free from spurious reflections off of the domain boundaries. 
The initial pressure profile is set by transforming the Friedlander profile shown in Eq.~(\ref{eq:averageFriedlanderEquation}), using $ t =  -\frac{x - d_2}{c_\mathrm{w}} $ and assuming the pre-shock pressure value to be equal to $ p_0 $: 
\begin{equation}
    p(x) = (p_{\mathrm{max}, 1} - p_0) \left(1 - \frac{d_2 - x}{c_\mathrm{w}\overline{t}_I} \right)\exp[{-\overline{\beta} (d_2 - x)/(c_\mathrm{w}\overline{t}_I)}] + p_0,
    \label{eq:initial_Pressure}
\end{equation}
where, $ p_0 = 101325\ \si{\pascal}$ and $ x \in [-\infty, d_2]$. The domain origin is located at the bottom-left corner, as illustrated in Fig.~\ref{fig:simulationSetup}(a). The term $ c_\mathrm{w} = 1456.2 \si{\meter\per\second} $ is the pre-shock sound speed of water and is needed to convert the pressure profile specified at a single spatial point into an initial pressure field defined over the whole domain. Once the pressure profile for the post-shock state is set, the initial condition for the velocity mixture and density are determined using the same normalized function:
\begin{align}
        \rho_\mathrm{w}(x) = (\rho_{\mathrm{max}, 1} - \rho_{0, w}) \left(1 - \frac{d_2 - x}{c_\mathrm{w}\overline{t}_I} \right)\exp[{-\overline{\beta} (d_2 - x)/(c_\mathrm{w}\overline{t}_I)}] + \rho_{0, w} \\
        u_{x, w}(x) = (u_{x, \mathrm{max}, 1} - u_{x, 0, w}) \left(1 - \frac{d_2 - x}{c_\mathrm{w}\overline{t}_I} \right)\exp[{-\overline{\beta} (d_2 - x)/(c_\mathrm{w}\overline{t}_I)}] + u_{x, 0, w},
        \label{eq:initial_density_speed}
\end{align}  
where the values for the peak density $ \rho_{\mathrm{max}, 1} $ and peak horizontal velocity $ u_{x, \mathrm{max}, 1} $ are deduced from the Rankine-Hugoniot jump conditions for the selected peak pressure. The values for $ \rho_{0, w} $ and $ u_{x, 0, w} $ are displayed in Table~\ref{tab:initialConditions}.

\begin{table}[h]
    \caption{Initial condition of the pre-shock state for water and PFH defined at $T_\mathrm{w} = T_\mathrm{P} = 298~\si{\kelvin}$.}
    \begin{tabular}{l|cc}
        \ & \textrm{Water} & \textrm{PFH} \\
        \hline
        $ p_0 $ [\si{\pascal}] & 101325 & 101325 \\
        $ \rho_0 $ [\si{\kilo\gram\per\meter\cubed}] & 1006.21 & 1680.1 \\
        $ u_{x, 0} $ [\si{\meter\per\second}] & 0 & 0 \\
        $ u_{y, 0} $ [\si{\meter\per\second}] & 0 & 0 \\
        \hline
    \end{tabular}
    \label{tab:initialConditions}
\end{table}

\subsection{Background-oriented schlieren (BOS) measurements}
\begin{figure}[ht]
    \centering
    \includegraphics[width=0.5\linewidth]{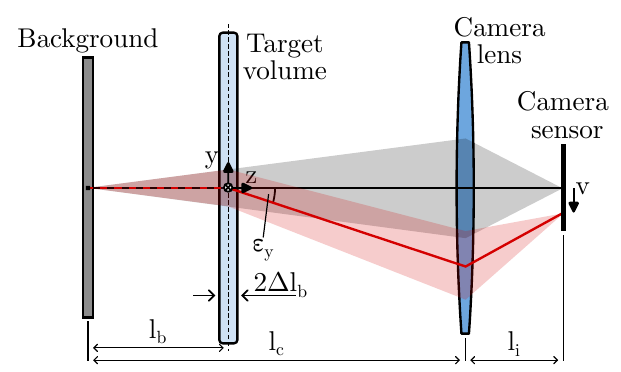}
    \caption{Schematic of a typical BOS setup. The deviation $ v $ of the background pattern, generated by the refraction of the light rays passing through the target volume, can be related to the deviation angle $ \epsilon_{\mathrm{y}} $ by geometrical consideration once the distances between the components $ l_b $, $ l_c $, and $ l_i $ are known.}
    \label{fig:BOS_basicSetup}
\end{figure}

In order to complement the numerical simulation and confirm the sign inversion of the shock wave across the droplet focal point, the background-oriented schlieren (BOS) method is employed since it is a non-invasive, quantitative optical measurement technique with the capability to measure the integrated density gradient through a  volume of interest~\cite{venkatakrishnan_density_2004, raffel_background-oriented_2015}.
A typical configuration for BOS technique measurements is shown in Fig.~\ref{fig:BOS_basicSetup}, and consists in both a target volume and a background positioned along a camera's optical axis, here schematized as a camera lens mounted in front of an image sensor. The camera's focal plane is set to be at the same position as the background. 
In the absence of the target volume, the light rays travel undisturbed from the background to the camera lens and are captured at the camera sensor. For example, light from the background center (solid line and gray shaded region) is collected at the center of the sensor, as illustrated in Fig.~\ref{fig:BOS_basicSetup}. The generated image is in this case referred to as the reference image.
In contrast, when the target volume is present, light rays starting from the background are refracted within the target volume and are generally focused at a different position on the sensor; the image thus generated is referred to as the distorted image. As an example, Fig.~\ref{fig:BOS_basicSetup} shows the cone of light traveling from the background center (red line and red shaded region) being distorted by the target volume before reaching the camera lens and focused on the camera sensor at a distance $ v $ from the reference.
The distortion of the background pattern obtained by comparing the reference image and the distorted image is defined as the displacement $ \bm{w} = [u,\ v] $ along the x- and y-direction. 
Here, the Fast Checker Demodulation (FCD) method \cite{wildeman_real-time_2018, shimazaki_background_2022} is employed to detect the displacements, which are proportional to the deviation angles $ \epsilon_{x} $, $ \epsilon_{y} $ between light rays in the presence and absence of the target volume (see Fig~\ref{fig:BOS_basicSetup}).
The relationship between the displacement and the deviation angle can be derived based on the near-field BOS formulation as: 
\begin{equation}
   \bm{w} = \frac{l_i}{l_b + l_c}\frac{(l_b + \Delta l_b)^2}{l_b}\bm{\epsilon} = M\frac{(l_b + \Delta l_b)^2}{l_b}\bm{\epsilon}
    \label{eq:BOS_displacement}
\end{equation}
where $ l_i $, $ l_b $, $ l_c $ are the distances between the different components of the optical setup,  $ M = l_{\mathrm{i}}/(l_\mathrm{b} + l_{\mathrm{c}}) $ is the camera magnification and $ \bm{\epsilon} = [\epsilon_x,\ \epsilon_y] $ is the vector of the deviation angles in the direction perpendicular to the optical axis \cite{van_hinsberg_density_2014}.
The near field correction for the standard BOS theory is here adopted because it is more suitable for measurements in which $l_b$ and $2\Delta l_b$ are of comparable magnitude.

The deviation angle can be expressed as the integral of the refractive index gradient along the optical axis, which is in turn related to the integrated density gradient via the Gladstone-Dale equation $ n_{\mathrm{P}} = K_{\mathrm{P}} \rho_{\mathrm{P}} + 1 $:
\begin{equation}
    \bm{\epsilon} 
    = [\epsilon_x,\ \epsilon_y] 
    = \frac{1}{n_{\mathrm{P}, 0}}\int_{l_b - \Delta l_b}^{l_b + \Delta l_b} \left[\frac{\partial n_{\mathrm{P}}}{\partial x}, \frac{\partial n_{\mathrm{P}}}{\partial y}\right] dz 
    = \frac{K_{\mathrm{P}}}{n_{\mathrm{P}, 0}}\int_{l_b - \Delta l_b}^{l_b + \Delta l_b} \left[\frac{\partial \rho_{\mathrm{P}}}{\partial x}, \frac{\partial \rho_{\mathrm{P}}}{\partial y}\right] dz, 
    \label{eq:BOS_deviationAngle}
\end{equation}
where $n_{\mathrm{P},0} = 1.25$ is the refractive index of PFH at ambient pressure, $ \rho_{\mathrm{P}} $ denotes its density, and $ K_{\mathrm{P}} = 1.49\times10^{-4}\ \si{\meter\cubed\per\kilogram} $ is the corresponding Gladstone–Dale constant. 

The density gradient field integrated along the camera line of sight can be reconstructed to obtain the density gradient distribution on a given $xy$-plane. In the present study, as in the work of \citet{yamamoto_contactless_2022} on the measurement of underwater shock waves in a microtube, the density gradient profile along the $z$-axis is assumed to be constant in order to simplify the calculation.
Under this assumption, the density gradient field on a specific $xy$-plane can be obtained by dividing the line-of-sight–integrated quantity by the thickness of the measurement region, $2 \Delta l_b$, yielding
\begin{equation}
\left[\overline{\frac{\partial \rho_{\mathrm{P}}}{\partial x}}, \overline{\frac{\partial \rho_{\mathrm{P}}}{\partial y}}\right]
=
\frac{n_{\mathrm{P}, 0}}{2 \Delta l_b K_{\mathrm{P}}}
\bm{\epsilon}.
\label{eq:BOS_densityGradient}
\end{equation}

By numerical integration of $ \overline{\partial \rho_{\mathrm{P}}}/\partial\mathrm{x} $ along the x-axis, the density function can be calculated:
\begin{equation}
    \rho_\mathrm{P}(x) = \rho_\mathrm{P, 0} + \int_{x_0}^{x} \overline{\frac{\partial \rho_\mathrm{P}}{\partial x^{\prime}}}\ dx^{\prime} 
    \label{eq:density_field}
\end{equation}
where $ \rho_\mathrm{P, 0} $ is the density at the initial integration point $ \mathrm{x}_0 $, which can be approximated by the reference value (see Table~\ref{tab:initialConditions}). The value of the integral is approximated with the composite trapezoidal rule. To ensure that the density at the two extremes of the integration domain are the same, as expected from a spatially limited traveling pressure pulse, the average value of the centerline signal is subtracted before integration. 

%


\begin{figure}[ht]
    \centering
    \includegraphics[width=\linewidth]{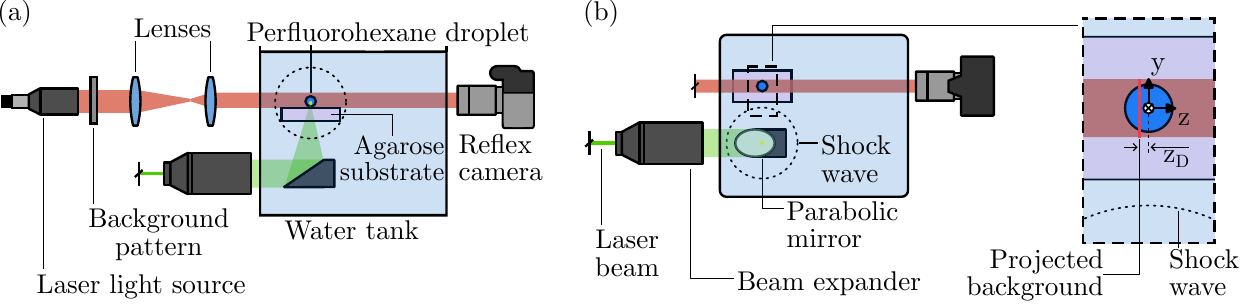}
    \caption{(a) Side and (b) top view schematic of the experimental setup used to perform density gradient measurements using the projected BOS technique. The shock wave is generated as described in Fig.~\ref{fig:experimentalSetupXRay}. A commercial single frame reflex camera combined with a pulsed laser illumination system is used to record the images with an effective exposure time of $ t_{\mathrm{e}} = 10\ \si{\nano\second} $. Two lenses are added in the light path to create a projection of the background inside the droplet, at a distance $ Z_{\mathrm{D}} $ from the droplet center plane, as shown in the detail view.}
    \label{fig:BOS_setup}
\end{figure}

The complete setup used for the BOS measurements is shown in Fig.~\ref{fig:BOS_setup}. 
Using two lenses, a projection of the background, displaying a checkered pattern, is created inside the droplet at a distance $ l_b \simeq0.16\ \si{\milli\meter} $ from its center plane, allowing non-invasive measurements of the density field. 
Shock waves are generated using a system very similar to the one displayed in Fig.~\ref{fig:experimentalSetupXRay}. The $ R \sim 1.0\ \si{\milli\meter} $ PFH droplet is at rest on a flat, acoustically transparent, 2\% agarose gel substrate to keep it in place during the interaction with the shock wave. A commercial reflex camera (Nikon D7200) is used to record single frames of the shock wave propagation inside the droplet. The exposure time is controlled by a pulsed diode laser light source (Cavitar CAVILUX Smart UHS) using a fixed pulse length of $ t_{\mathrm{e}} = 10\ \si{\nano\second} $.    
Once the Nd:YAG laser is fired, a TTL pulse is sent to a delay generator, which triggers the laser illumination system applying a user-set delay $ \Delta t_{\mathrm{e}} $, following a similar approach to the one described in detail by \citet{ichihara_high-resolution_2025}. The experiment is repeated by varying $ \Delta t_{\mathrm{e}} $ between 4.5 \si{\micro\second} and 11.0 \si{\micro\second} to capture different instants of the shock wave transmission. The incoming shock wave peak pressure is set to a value slightly lower than the one selected for the x-ray experiments ($ p_{\mathrm{max}, 1} = 14.39 $ \si{\mega\pascal}). This precaution prevents the inception of cavitation and allows the use of the same droplet for multiple measurements.

\begin{figure}[h]
    \centering
    \includegraphics[width=0.5\linewidth]{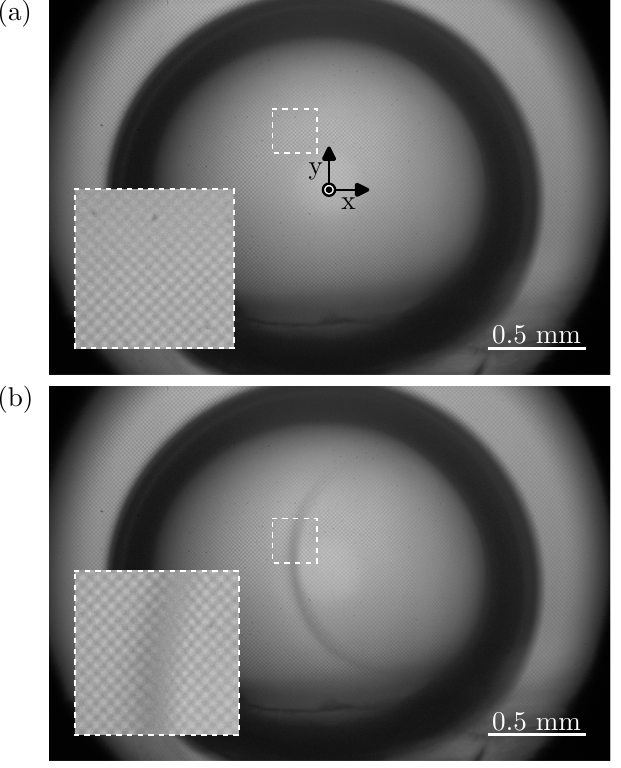}
    \caption{(a) Reference image of the projected background inside the PFH droplet. (b) Image of the distorted projected background due to the shock wave traveling inside the droplet. The dark area close to the edges of the drop are generated by the refraction of the light rays due to the curved interface and the refraction index mismatch. An enlarged view of the area inside the dashed-border rectangle is shown in both images at their bottom-left corner.}
    \label{fig:BOS_rawData}
\end{figure}

An example of a reference and distorted images captured during a BOS measurement is shown in Fig.~\ref{fig:BOS_rawData}(a) and Fig.~\ref{fig:BOS_rawData}(b), respectively. 
The distortion of the checkered pattern, caused by the passage of the shock wave, is visible at the center of the droplet in Fig.~\ref{fig:BOS_rawData}(b). 
The presence of a spherical droplet introduces an additional index of refraction mismatch that causes a dark region at the droplet edge and additional distortion around the droplet center, the latter assumed to be negligible in this study.
A quantitative evaluation of pressure waves inside the droplet that accounts for the additional distortion is presented in the work of \citet{ichihara_measurement_2026}. 

\subsection{Radial dynamics of air-filled cavitation nuclei}
To investigate the hypothetical presence of air cavities within the perfluorocarbon droplet, the radial dynamics of multiple nanometric bubbles is simulated using the Keller-Miksis equation \cite{keller_bubble_1980} in a slightly modified form, as employed in~\citet{kamath_numerical_1989} and described in ~\citet{prosperetti_bubble_1986}. 
The Keller--Miksis model is chosen over the Rayleigh--Plesset model because of its suitability in describing large amplitude oscillations.
The equation reads:  
\begin{equation}
    \begin{split}
    \left( 1 -  \frac{\dot{R}_\mathrm{b}}{c_{\mathrm{P}}} \right) R_\mathrm{b} \ddot{R}_\mathrm{b} + \frac{3}{2} \left ( 1 - \frac{\dot{R}_\mathrm{b}}{3c_{\mathrm{P}}} \right) \dot{R}^2_\mathrm{b} = \frac{1}{\rho_{\mathrm{P}}}\left( 1 + \frac{\dot{R}_\mathrm{b}}{c_{\mathrm{P}}}\right)\left[p(R_\mathrm{b}, t) - p_{\infty}\left( t + \frac{R_\mathrm{b}}{c_{\mathrm{P}}}\right) \right] + \frac{R_\mathrm{b}}{c_{\mathrm{P}}}\frac{d p(R_\mathrm{b}, t)}{dt},
    \end{split}
    \label{eq:modified_Keller_Miksis}
\end{equation}
where $ R_\mathrm{b},\ \dot{R}_\mathrm{b} $ are the bubble radius and its time derivative, $ c_{\mathrm{P}} $ is the sound speed of PFH, $ \rho_{\mathrm{P}} $ its density. The term $ p_{\infty}(t) $ is the pressure field at a specific position (x,y) inside the droplet and is extracted from the ECOGEN simulation.
The bubble is assumed to be filled with air and PFH vapor at the saturation pressure. The initial radius is set to $ R_{_\mathrm{b},0} = 3\ \si{\nano\meter} $, which corresponds closely to the average nucleus size measured in water in the work by~\citet{mancia_acoustic_2021}.
The pressure inside the bubble is assumed to be uniform, and the gas transformation is modeled as isothermal. The pressure of the liquid at the bubble boundary $ p(R_\mathrm{b}, t) $ can then be modeled as: 
\begin{equation}
    p(R_\mathrm{b}, t) = p_{\mathrm{g},0}\left(\frac{R_{\mathrm{b}, 0}}{R_\mathrm{b}} \right)^{3} - \frac{2\sigma_{\mathrm{air-P}}}{R_\mathrm{b}}
    + p_\mathrm{v} - 4\mu_{\mathrm{P}}\frac{\dot{R}_\mathrm{b}}{R_\mathrm{b}},  
\label{eq:modified_Keller_Miksis_Pressure_model}
\end{equation}
where $ p_{\mathrm{g},0} = p_{\infty}(t = 0) + \frac{2\sigma_{\mathrm{air-P}}}{R_{\mathrm{b}, 0}} - p_{\mathrm{v}}$ is the gas equilibrium pressure, $ \sigma_{\mathrm{air-P}} $ is the surface tension between air and liquid PFH, $ p_{\mathrm{v}} $ is the PFH vapor partial pressure and $ \mu_{\mathrm{P}} $ is the liquid PFH's viscosity. The numerical values of the parameters used are reported in Table~\ref{tab:modified_Keller_Miksis_Parameter}.
Variations in the internal temperature field, predictable using a more sophisticated model \cite{prosperetti_nonlinear_1988}, are neglected to reduce computational time.  

\begin{table}[h]
    \centering
    \caption{Numerical values of the parameter used to estimate the radial dynamics of air-filled nanobubbles in the PFH droplet, driven by the pressure obtained from the numerical simulation.}
    \begin{tabular}{l|c}
        Parameter & Value \\ 
        \hline
        $ R_{\mathrm{b}, 0}\ [\si{\nano\meter}] $ & 3  \\
        $ \dot{R}_{\mathrm{b}, 0}\ [\si{\meter\per\second}] $ & 0  \\
        $ c_{\mathrm{P}}\ [\si{\meter\per\second}] $ & 485  \\
        $ \rho_{\mathrm{P}}\ [\si{\kilogram\per\meter\cubed}] $ & 1680.1  \\
        $ \sigma_{\mathrm{air-P}}\ [\si{\newton\per\meter}] $ & $ 12 \times 10^{-3} $ \\
        $ p_{\mathrm{v}}\ [\si{\pascal}] $ & $ 31495 $ \\
        $ \mu_{\mathrm{P}}\ [\si{\pascal\second}] $ &  $ 0.64 \times 10^{-3} $ \\
    \end{tabular} 
    \label{tab:modified_Keller_Miksis_Parameter}
\end{table}

\subsection{Critical nuclei formation rate estimation using Classical Nucleation Theory}

In the homogeneous nucleation model, vapor nuclei (agglomerates of molecules) of various sizes are continuously created and varying in size due to the random thermal motion of PFH liquid molecules. 
When the pressure of a liquid at a fixed temperature rises above its saturation pressure, the presence of a vapor nucleus of any size increases the Gibbs Free Energy of the liquid-vapour system, ultimately leading to the spontaneous disappearance of the nucleus. However, if the pressure drops below the saturation pressure, the Gibbs Free Energy function presents a maximum for a vapor-filled nucleus of radius $ R_{\mathrm{n}} = R_\mathrm{c} $, with $ R_\mathrm{c} $ being a critical nucleus size or critical radius. The presence of a local maximum in the Free Energy of the systems means that a vapor nucleus with a radius $ R_{\mathrm{n}} < R_\mathrm{c} $ will spontaneously disappear, while one with $ R_{\mathrm{n}} > R_\mathrm{c} $ will grow indefinitely, eventually converting the entire liquid phase into vapor. The presence of an energy barrier at a critical nucleus size is the physical phenomenon that allows the existence of superheated fluids, which are in a condition of metastable thermodynamical equilibrium. 
As the pressure decreases further, both the critical radius and the energy barrier progressively lower and vanish when the thermodynamic stability limit of the system with respect to small fluctuations, known as spinodal limit, is reached. 


The expression for the rate of formation of critical nuclei per unit volume $ J(T_{\mathrm{P}}, p_{\mathrm{P}}) $, originally derived using Classical Nucleation Theory (CNT)~\cite{carey_liquid-vapor_2020} and recently updated to account for the energy barrier vanishing at the spinodal by Qin~\emph{et al.} \cite{qin_Predicting_2021}, reads as follows:
\begin{equation}
    J[T_{\mathrm{P}}, p_{\mathrm{P}}(x, y, t)] = J_0\exp \left({\frac{-16\sigma_r(T_{\mathrm{P}}, p_{\mathrm{P}})^3}{3k_{\mathrm{B}} T_{\mathrm{P}}[p_{\mathrm{v}}(T_{\mathrm{P}}, p_{\mathrm{P}}) - p_{\mathrm{P}}]^2}}\right),
    \label{eq:HomogenousNucleation_J}
\end{equation}
where $ T_{\mathrm{P}}$ is the droplet temperatur and  $ k_{\mathrm{B}} $ is the Boltzmann constant. The term $ p_{\mathrm{P}}(x, y, t) $ is the instantaneous pressure at a specific location $ (x, y) $ inside the droplet with the addition of the Laplace pressure contribution $ 2 \sigma_{\mathrm{w}-\mathrm{P}}/R $ due the curved droplet interface, with $ \sigma_{\mathrm{w}-\mathrm{P}} $ being the water-PFH interfacial tension \cite{nishikido_interfacial_1989} and $ R $ the droplet radius. 
Here, $ \sigma_r(T_{\mathrm{P}}, p_{\mathrm{P}}(x, y, t)) $ is the surface tension between the PFH liquid and vapor in a nucleus of critical size:
    \begin{equation}
    \begin{split}
        \sigma_r[T_{\mathrm{P}}, p_{\mathrm{P}}(x, y, t)] =\sigma_{\infty}[T_{\mathrm{P}}(x, y, t)] \left(1 - \frac{p_{\mathrm{sat, P}}(T_{\mathrm{P}}) - p_{\mathrm{P}}(x, y, t)}{p_{\mathrm{sat, P}}(T_{\mathrm{P}}) - p_{\mathrm{spin, P}}(T_{\mathrm{P}})}\right)\left(1 + 0.5\frac{p_{\mathrm{sat, P}}(T_{\mathrm{P}}) - p_{\mathrm{P}}(x, y, t)}{p_{\mathrm{sat, P}}(T_{\mathrm{P}}) - p_{\mathrm{spin, P}}(T_{\mathrm{P}})}\right)^2,
    \end{split}
    \label{eq:sigma_critical_nucleus}
\end{equation}
where $ \sigma_{\infty}(T_{\mathrm{P}}) $ is the macroscopic surface tension of the flat PFH liquid-vapor interface and $ p_{\mathrm{sat, P}} $ is the saturation pressure at the temperature $ T_{\mathrm{P}} $ of the fluid calculated using the Antoine equation with PFH-specific constants~\cite{crowder_vapor_1967}. The spinodal pressure of PFH $ p_{\mathrm{spin, P}} $ is calculated using the Redlich-Kwong Equation of State~\cite{poling_Properties_2020} and the critical properties reported in~\citet{gao_equations_2021}. The term $ p_{\mathrm{v}}(T_{\mathrm{P}}, p_{\mathrm{P}}) $ is the vapor pressure inside a critical nucleus, calculated as:
\begin{equation}
    p_{\mathrm{v}}(T_{\mathrm{P}}, p_{\mathrm{P}}) = p_{\mathrm{sat, P}}(T_{\mathrm{P}})\exp\left({\frac{p_{\mathrm{P}}(T_{\mathrm{P}}) - p_{\mathrm{sat, P}}(T_{\mathrm{P}})}{\rho_\mathrm{P}R_{\mathrm{m}} T_{\mathrm{P}}}}\right),
    \label{eq:nucleus_vapor_Pressure}
\end{equation}
with $ R_{\mathrm{m}} $ being the specific gas constant for PFH. The pre-exponential factor representing the maximum value attainable for the nucleation rate is $ J_0 = \sqrt{3\sigma_{\infty}\rho_{\mathrm{P}}/\pi m^3} $, where $ m $ is the molecular mass of PFH. The values of the thermodynamic parameters defined above are available in Table~\ref{tab:MCNT_Parameter}.

\begin{table}[]
    \centering
    \caption{Numerical values of the thermodynamic parameters used to estimate the number of critical nuclei generated during the shock wave interaction with the PFH droplet. The value for the saturation $ P_{\mathrm{sat, P}} $ and spinodal pressure are reported $ P_{\mathrm{spin, P}} $ only at the temperature considered in the calculation.}
    \begin{tabular}{l|c}
        Parameter  & Value \\
        \hline
        $ k_{\mathrm{B}} [\si{\meter\squared\kilogram\per\second\squared\per\kelvin}]$ & $ 1.38\times10^{-23} $ \\
        $ \sigma_{\infty}\ [\si{\newton\per\meter}] $  & $ 12\times10^{-3} $ \\
        $ \sigma_{\mathrm{w}-\mathrm{P}}\ [\si{\newton\per\meter}] $  & $ 49\times10^{-3} $ \\
        $ T_{\mathrm{P}}\ [\si{\kelvin}] $ & 298 \\
        $ P_{\mathrm{sat, P}}(T_{\mathrm{P}}) [\si{\pascal}]$ & $ 2.93\times10^4 $ \\
        $ P_{\mathrm{spin, P}}(T_{\mathrm{P}}) [\si{\pascal}]$ & $- 7.43\times10^{6}$ \\
        $ \rho_{\mathrm{P}}\ [\si{\kilogram\per\meter\cubed}] $ & 1680.1 \\
        $ R_\mathrm{m} $ [\si{\joule\per\kilogram\per\kelvin}] & 24.6 \\ 
        $ m\ [\si{\kilogram}] $ & $ 5.61\times10^{-22} $ \\

    \end{tabular}
        \label{tab:MCNT_Parameter}
\end{table}

\section{Results \& Discussion}
\subsection{Cavitation detection using high-speed x-ray imaging}
\label{sec:xRayResults}

\begin{figure}[h]
    \centering
    \includegraphics[width = \linewidth]{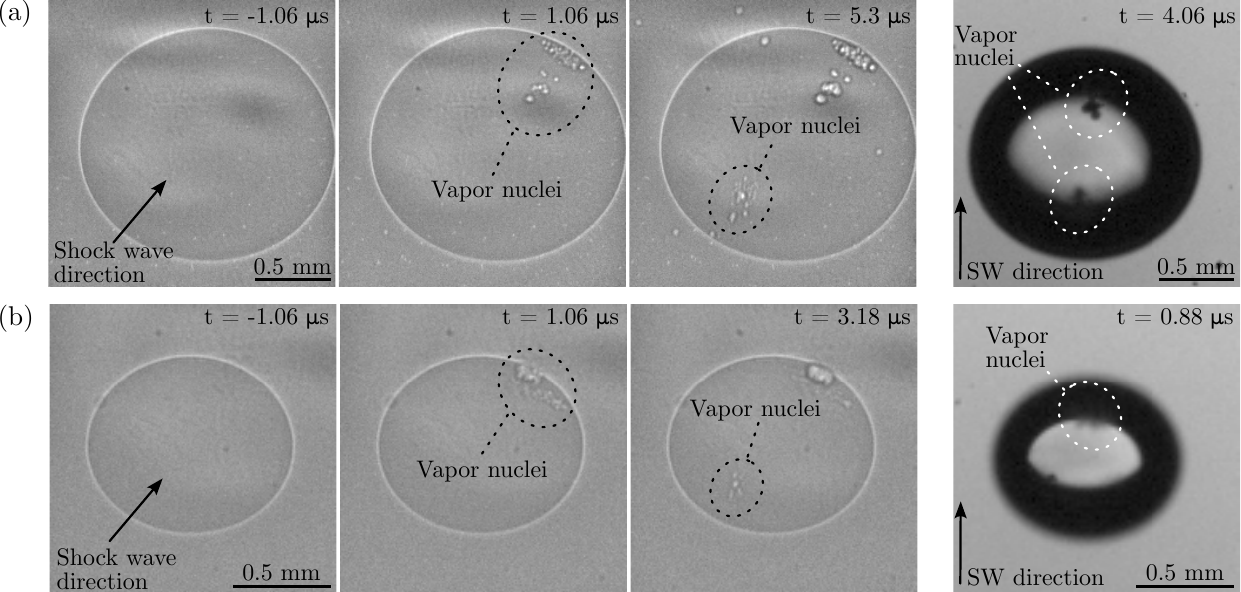}
    \caption{X-ray phase-contrast (left) and shadowgraph (right most) snapshots of the interaction between a shock wave traveling in water and a PFH droplet of radius (a)  $ R = 825\ \si{\micro\meter} $ and (b) $ R = 470\ \si{\micro\meter} $. The shock wave direction is indicated by an arrow, and the regions in which cavitation activity is detected are circled with a dashed line. The time $ t = 0\ \si{\micro\second}$ here corresponds to the time at which nucleation is first detected in the radiographs.}
    \label{fig:experimentalXRaySnapshots}
\end{figure}

Fig.~\ref{fig:experimentalXRaySnapshots} shows a selection of snapshots from the x-ray phase-contrast recordings and a single frame from the corresponding shadowgraph. The time origin ($ t = 0 $) corresponds to the instant of the first detected nucleation event. 
In all of the snapshots presented, the direction in which the shock wave travels is indicated by an arrow. The recordings are available as Supplementary Material (see Appendix~\ref{sec:suppMat} for details on the content). 

The radiographs suggest two distinct nucleation patterns: in the first case in Fig.~\ref{fig:experimentalXRaySnapshots}(a), pronounced cavitation activity is detected in the conical region located at the distal (top-right) side of the droplet with respect to the incoming wave. Interestingly, nucleation is more pronounced close to the droplet interface. A second nucleation area appears later at the proximal (bottom-left) side of the droplet. In the second case in Fig.~\ref{fig:experimentalXRaySnapshots}(b), cavitation activity at the distal side is limited to the region close to the interface, while at the proximal side nucleation can be observed approximately at the same location as in Fig.~\ref{fig:experimentalXRaySnapshots}(a). Due to its relatively large size, the droplet in free fall assumes a pseudo-elliptical shape caused by the inability of the interfacial tension forces to stabilize the interface against inertial and hydrodynamic forces. The droplet radii in Fig.~\ref{fig:experimentalXRaySnapshots} are estimated by fitting an ellipse on the droplet edge, extracting the semi-major and -minor axes and taking their geometrical mean. The equivalent radius corresponds to 825 \si{\micro\meter} and 470 \si{\micro\meter} for case (a) and (b), respectively.

\begin{figure}
    \centering
    \includegraphics[width = \linewidth]{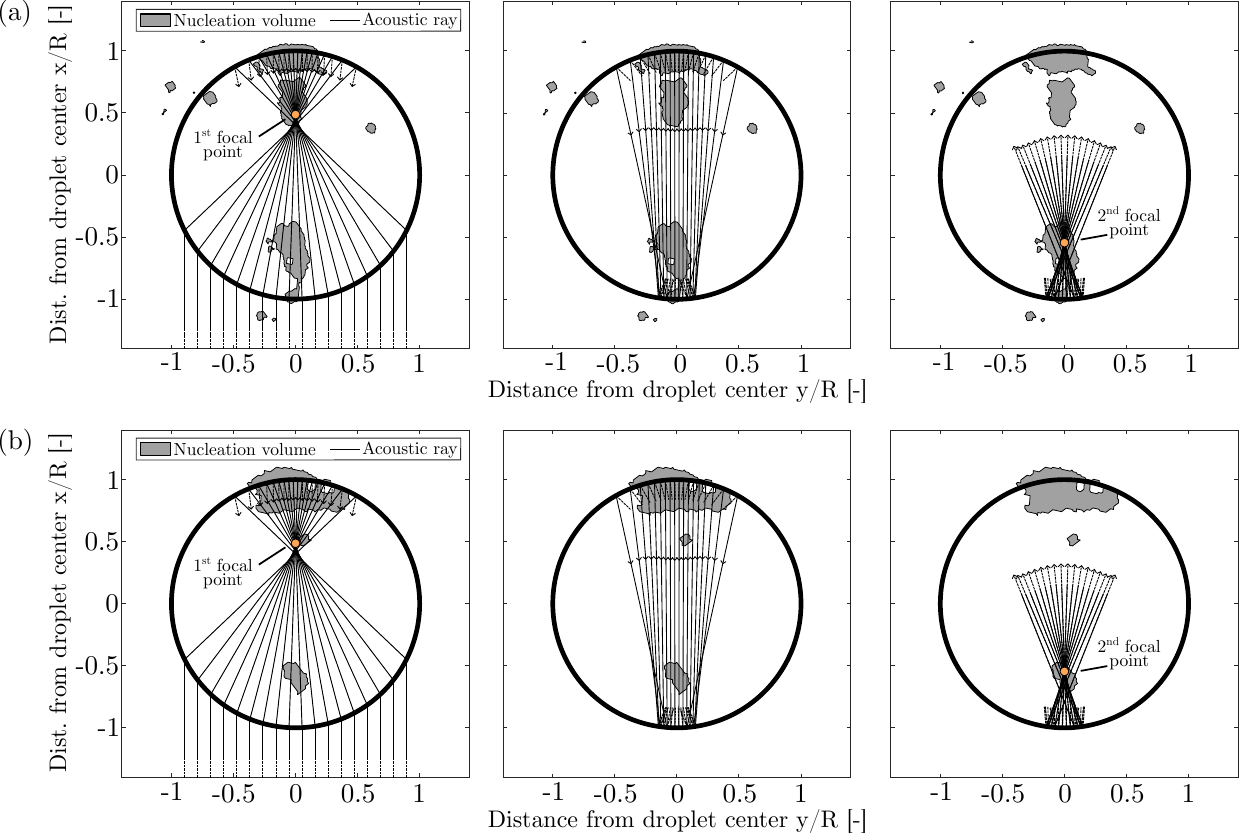}
    \caption{Trajectory followed by parallel acoustic rays interacting with a spherical PFH droplet surrounded by water. The acoustic path is shown from the first refraction at the proximal side to the first reflection on the distal side (left column), from the first reflection to the second reflection at the proximal side (center column), and from the second reflection up to reaching a second focal point on the proximal side (right column). The red dot on the leftmost plot represents the focal point calculated from Eq.~(\ref{eq:focalPointEq}). The light blue shaded areas in (a)-(b) represent the regions where nucleation activity is detected in a droplet of radius $ R = 825\ \si{\micro\meter} $ and (b) $ R = 470\ \si{\micro\meter}$ respectively, obtained from the videos in Fig.~\ref{fig:experimentalXRaySnapshots}(a)-(b). The x- and y-distances have been made dimensionless by fitting an ellipse over the droplet edge and normalizing them by the value of the semi-axis for better comparison.}
    \label{fig:rayAcoustics}
\end{figure}

The interaction of a shock wave with a liquid droplet can be complex, involving several rounds of reflections inside the droplet core. The geometrical scattering approximation~\cite{escoffre_therapeutic_2016}, combined with the experimentally measured nucleation maps, can be exploited to give an intuitive explanation of the phenomenon. Fig.~\ref{fig:rayAcoustics}(a) and (b) show the acoustic ray trajectories overlapped with the experimental nucleation regions detected from the x-ray phase contrast videos depicted in Fig.~\ref{fig:experimentalXRaySnapshots}(a) and (b), respectively. The presence of bubbles in each x-ray snapshot is detected by subtraction of a reference image and application of an edge detection algorithm. The nucleation regions are then obtained by summation of all single-frame maps, identifying locations presenting cavitation activity throughout the whole recording. 
The ray path is estimated by considering both media (water and PFH) to be homogeneous and by assuming the incoming shock wave to be planar. The planar wave fronts propagate along a straight line in the bulk of the two media, while the change in direction at the droplet interface is governed by the acoustic equivalent of Snell's law \cite{kinsler_fundamentals_2000}:
 \begin{equation}
      \frac{\sin(\theta_i)}{c_i} =  \frac{\sin(\theta_t)}{c_t},
     \label{snellLawAcoustic}
 \end{equation}
where $ \theta_i $ and $ \theta_t $ are the incident and transmission angles of the wave with respect to the vector normal to the local interface, and $ c_i $ and $ c_t $ the speeds of sound in the medium before and after the boundary.
Due to the lower sound speed of the perfluorocarbon inside the droplet with respect to the surrounding water, the incoming rays are bent towards the center of the drop and focus close to its distal side (see Fig.~\ref{fig:rayAcoustics}(a)-(b), left column) at the position:
\begin{equation}
    f = R\left(\frac{c_i}{c_i - c_t} - 1\right),
    \label{eq:focalPointEq}
\end{equation}
where $f$ is the distance between the focal point and the droplet center along the shock wave propagation direction and $R$ is the equivalent droplet radius. Assuming a speed of sound of $ c_i = 1481 $ \si{\meter\per\second} and $ c_t = 485 $ \si{\meter\per\second} \cite{prasanna_shock_2025} for water and PFH, the focal point is located at a distance $ f = 0.49R $ from the droplet center. After the wave crosses the focal point, it reflects from the distal interface and starts traveling towards the proximal interface (Fig.~\ref{fig:rayAcoustics}(a)-(b), center column) where yet another reflection occurs, and the wave is refocused at the proximal side (right column). 

Interestingly, in the case of Fig.~\ref{fig:rayAcoustics}(a), a nucleation region appears right after the predicted geometrical focus. Although shock waves generated by commercial lithotripters are typically followed by a strong tensile tail that, if focused, can generate the negative pressure necessary to induce nucleation through amplification \cite{wess_fragmentation_2020}, the shock waves generated here through optical breakdown exhibit practically no negative pressure, as seen in the inset of Fig.~\ref{fig:experimentalSetupXRay}(b). Therefore, the tensions responsible for the observed cavitation cannot be attributed to the amplification of negative pressure of the incoming wave.

A more convincing interpretation for cavitation inception inside the droplets can be given by the occurrence of Gouy phase shift of the focusing shock wave. The phase shift that occurs while crossing the focus can generate negative pressure, justifying the nucleation of bubbles experimentally observed at the proximal and distal side. Moreover, it provides an intuitive explanation for the appearance of a nucleation region at the interface of the droplet's distal side, as visualized in both Fig.~\ref{fig:rayAcoustics}(a) and (b). After tension generation at the focal point, reflection at the distal interface takes place. Since PFH has a lower acoustic impedance $ Z_{\mathrm{P}} = \rho_\mathrm{P} c_\mathrm{P} \sim 8 \times 10^{5} $ than the surrounding water $ Z_\mathrm{w} = \rho_\mathrm{w} c_\mathrm{w} \sim 15 \times 10^{5} $, waves traveling in PFH and reflected at the PFH-water boundary maintain the same sign~\cite{kinsler_fundamentals_2000}. Therefore, the reflected wave will interfere constructively with the one still traveling towards the boundary, generating a negative pressure region which can explain the enhanced nucleation activity.

\subsection{Numerical simulations of the pressure field}
\label{sec:numericalResults}

\begin{figure}[ht!]
    \centering
    \includegraphics[width = 0.99\linewidth]{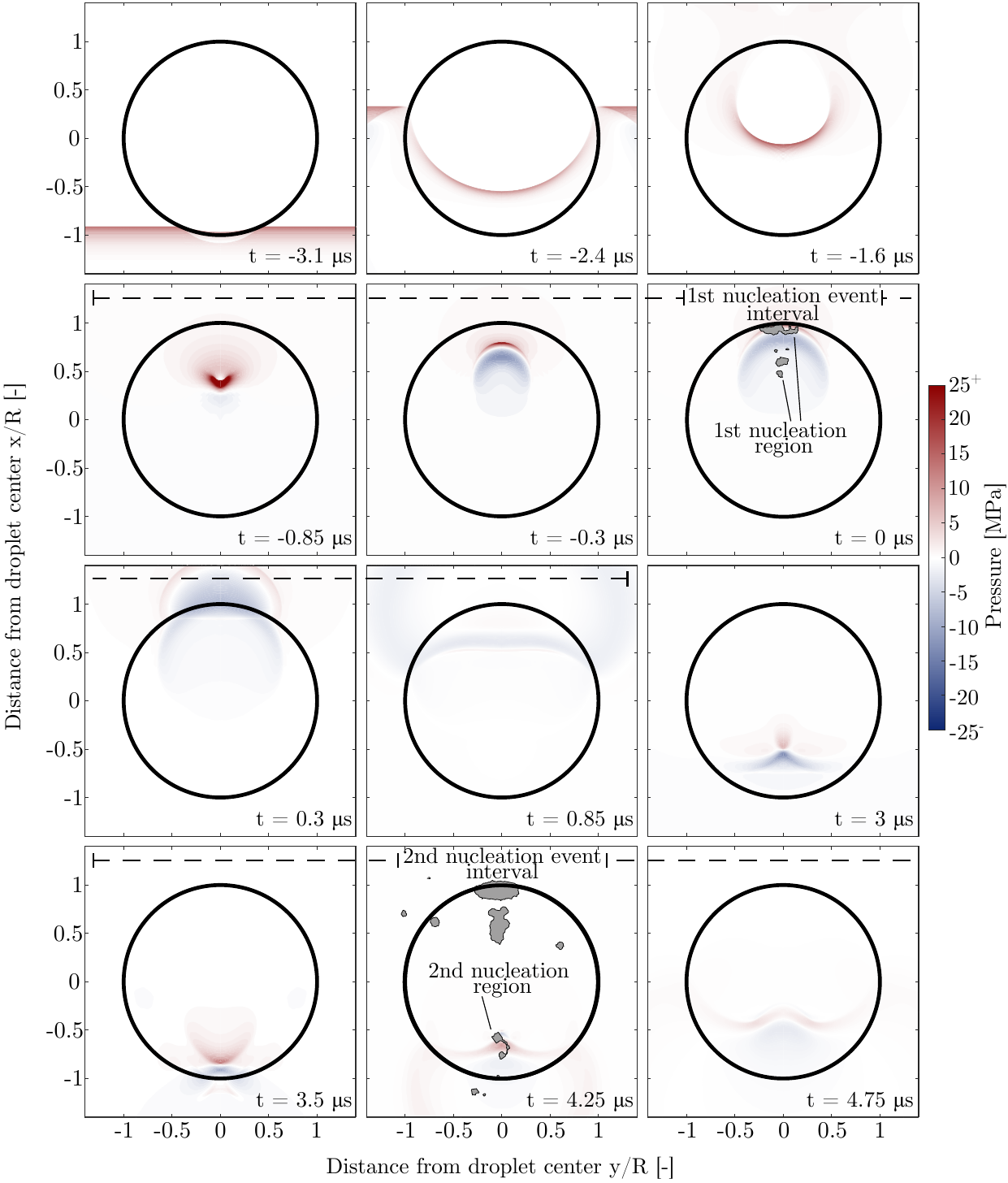}
    \caption{Numerical simulation of the pressure field generated by the interaction of a Friedlander-type shock wave (Eq.~(\ref{eq:friedlanderEquation}), peak positive pressure $ p_{\mathrm{max}, 1} = 14.39\ \si{\mega\pascal} $) with an 850-\si{\micro\meter}-radius PFH droplet. The spatial coordinates have been normalized by the droplet radius. The pressure scale has been limited between the values $ -25\ \si{\mega\pascal} $ and $ +25\ \si{\mega\pascal} $ to better visualize low pressure values. In particular, the pressure field at time $ t = -0.85~\si{\micro\second} $ presents a saturated color map. The full range of pressure obtained in the simulation lies between 247 \si{\mega\pascal} and -17 \si{\mega\pascal}. The gray areas superimposed to the pressure field represents the experimentally detected nucleation region at a specific timestep. The uncertainty range $ \pm R/(2c_{\mathrm{P}}) $ for the time instant in which nucleation appears in the recordings is reported with a dashed line. Here, $ t = 0\ \si{\micro\second} $ corresponds to the time at which the first cavitation event is detected, while the time occurrence of the second nucleation event is calculated from the experimental recordings.} 
    \label{fig:00XSimulation}
\end{figure}

\begin{figure}[ht!]
    \centering
    \includegraphics[width = 0.99\linewidth]{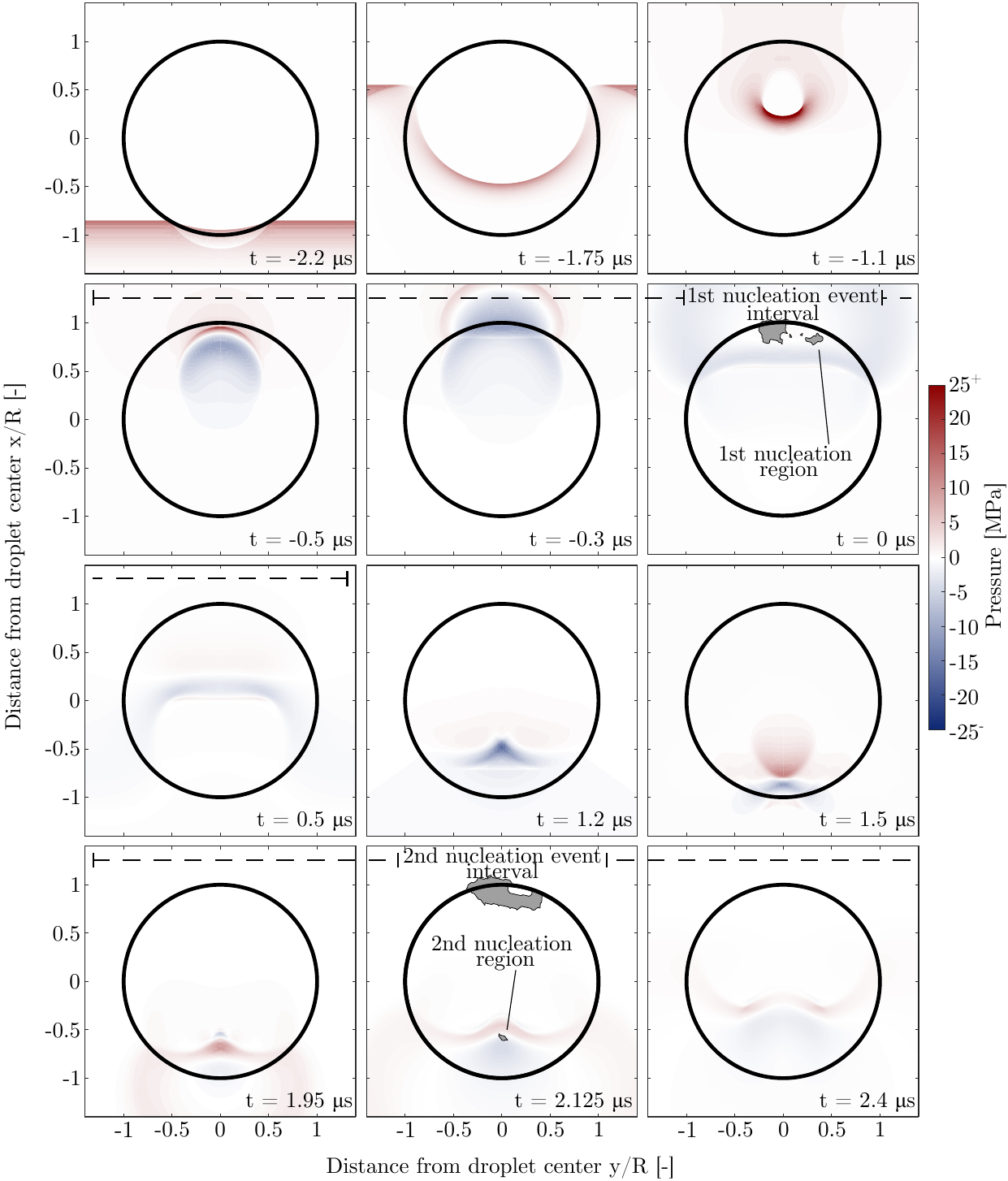}
    \caption{Numerical simulation of the pressure field generated by the interaction of a Friedlander-type shock wave (Eq.~(\ref{eq:friedlanderEquation}), peak positive pressure $ p_{\mathrm{max}, 1} = 14.39\ \si{\mega\pascal} $) with a 470-\si{\micro\meter}-radius PFH droplet. The spatial coordinates have been normalized by the droplet radius. The pressure range has been limited between the values $ -25\ \si{\mega\pascal} $ and $ +25\ \si{\mega\pascal} $ to better visualize low pressure values. In particular, the pressure field at time $ t = -1.1\ \si{\micro\second} $ presents a saturated color map. The full range of pressure obtained in the simulation lies between 303 \si{\mega\pascal} and -17 \si{\mega\pascal}. The gray areas superimposed to the pressure field represents the experimentally detected nucleation region at the correspondent timestep. The uncertainty range $ \pm R/(2c_{\mathrm{P}}) $ for the time instant in which nucleation appears in the recordings is reported with a dashed line. Here, $ t = 0\ \si{\micro\second} $ corresponds to the time at which the first cavitation event is detected, while the time occurrence of the second nucleation event is calculated from the experimental recordings.}
    \label{fig:0006Simulation}
\end{figure}

To further investigate tension generation upon focusing of the shock wave, 2D-axisymmetric numerical simulation of the interaction between a planar shock wave in water and a spherical PFH droplet are performed. Both Fig.~\ref{fig:00XSimulation} and Fig.~\ref{fig:0006Simulation} show the pressure field inside the droplet generated by focusing a shock wave of peak positive pressure $ p_{\mathrm{max, 1}} = 14.39\ \si{\mega\pascal} $, estimated from the experimental characterization of the setup (see Section~\ref{sec:xrayMethods}). The complete animations are available as Supplementary Material (see Appendix~\ref{sec:suppMat} for further information). The shadowgraph video is used to synchronize the x-ray recordings with the numerical simulation. When the wavefront is visible within the proximal side of the droplet ($ x/R < 0 $ in Fig.~\ref{fig:00XSimulation} and \ref{fig:0006Simulation}), its position can be estimated as $ \bar{x} =  -R/2\ \pm R/2 $, with the origin placed at the droplet center. 
The large uncertainty is due to the uncertainty of the angle between the wave front speed perpendicular to the droplet interface and the optical axis of the camera. The corresponding  time instant $ t =  \overline{t}_{R/2} $ is extracted from the video recordings. The time of arrival $ \overline{t}_0 $ of the wave front at the droplet interface can then be estimated by subtracting the time the shock wave needs to travel half the radius from $ \overline{t}_{R/2} $:
\begin{equation}
    \overline{t}_0 = \overline{t}_{R/2} - R/(2c_\mathrm{P})\ \pm\ R/(2c_\mathrm{P}), 
    \label{eq:xRaySimSync}
\end{equation}
where the shock wave speed is approximated using the reference sound speed of PFH, $ c_\mathrm{P} = 485\ \si{\meter\per\second} $. The time of arrival is then subtracted from the experimental time at which the first nucleation event occurs in the recordings [the same value used to define the time origin in Figs~\ref{fig:experimentalXRaySnapshots}(a)-(b)], and the results is used to shift the time origin of the numerical simulation (which start, with the wavefront located at the interface), therefore obtaining a common time scale between experimental and numerical results.

In the case of Fig.~\ref{fig:00XSimulation}, the droplet radius is $ R = 850\ \si{\micro\meter} $ to model the dynamics observed in Fig.~\ref{fig:experimentalXRaySnapshots}(a). The time $ t = 0\ \si{\micro\second} $ corresponds to the instant at which nucleation is detected in the related experiments. The numerical results clearly show the shock wave focusing inside the droplet ($ -3.1\ \si{\micro\second} < t < -0.85\ \si{\micro\second} $), with the focal point located at a position close to $ f_s = 0.5 R $, which is in good agreement with the theoretical prediction from Eq.~(\ref{eq:focalPointEq}). Upon crossing of the focal point, a strong negative pressure peak is generated and then transmitted in the droplet bulk. At the distal interface, the traveling shock wave is reflected at the PFH-water boundary while keeping its negative sign, therefore interfering constructively with the incoming wavefront and generating a strong tension region close to the droplet edge. The first nucleation event, whose interval is indicated in Fig.~\ref{fig:00XSimulation} with a dashed line centered at $ t = 0\ \si{\micro\second} $, occurs during the time span in which the wave is focused, generates a strong negative pressure at the focus, and is reflected at the distal interface. After the reflection, the newly created negative pressure wavefront travels towards the proximal side of the droplet. Here, as can be seen from the ray trajectory in Fig.~\ref{fig:rayAcoustics}, some rays are already crossing each other's path (shown in the central graph, around $ x/R \sim -0.5 $). This further increases the value of the pressure and creates two curved wave fronts in the left and right region of the droplet, as seen in Fig.~\ref{fig:00XSimulation} at the time $ t = 3\ \si{\micro\second}$. Once these two wave fronts meet at the droplet centerline, they generate a strong negative pressure region, in which cavitation is likely to occur. Immediately after ($ t = 3.5\ \si{\micro\second}$), a compression wave is generated, likely due to an effect similar to the Gouy phase shift at $ x/R \sim -0.5 $. However, nucleation is detected after the passage of both the negative and positive pressure wavefronts within the drop's distal side. Although this delay might be due to experimental uncertainties, in particular regarding the droplet radius, which can propagate inaccuracies to the later stages of the simulation, the time lag could also be caused by the strong compression wave that follows the negative pressure peak ($ t = 3.5\ \si{\micro\second} $), which might temporarily collapse the newly formed nuclei, making them only visible after the compression wave has passed ($ t > 4.25\ \si{\micro\second} $). 


Fig.~\ref{fig:0006Simulation} shows the interaction of a $ p_{\mathrm{max}, 1} = 14.39\ \si{\mega\pascal} $ shock wave with a PFH droplet of radius $ R = 470\ \si{\micro\meter} $, parameters chosen to reproduce the experimental results in Fig.~\ref{fig:experimentalXRaySnapshots}(b). Qualitatively, the pressure field appears very similar to the case of a larger droplet radius. Firstly, the shock wave focuses inside the droplet, reaching its maximum value at a position $ x/R \sim  0.5 $ from the droplet center ($ -2.2~\si{\micro \second} < t < -1.1~\si{\micro \second} $). Upon crossing of the focal point, negative pressure is generated ($ t \sim -0.5~\si{\micro\second} $) and the wave front is reflected at the distal PFH-water interface ($ t \sim 0.3~\si{\micro\second} $). Around $ t = 1.2\ \si{\micro\second} $, the two focused negative wave fronts meet, generating another high tension region at the proximal side.
In this case, the region around the focal point does not nucleate. A possible explanation could be that, for smaller droplets, the lensing effect is weaker and the pressure (both positive and negative) at the focus is lower. Additionally, progressive deterioration of the laser alignment, possibly occurring despite the conduction of periodic realignment of the optics between measurements, could lead to the generation of weaker and wider shock waves, a concurring effect preventing cavitation from happening.
Interestingly, cavitation is still triggered at the distal side ($ x/R \sim 1 $) and at the second focal point located at the position $ x/R \sim -0.5 $. This evidence suggests that tension generation is quite robust to degradation of the shape of the pressure pulse, an important feature to be considered for practical applications. 
Also in this case, cavitation at the proximal side does not seem to happen right after the negative pressure reaches its peak ($ t = 1.2\ \si{\micro\second} $), but around 1~\si{\micro\second} later, possibly due to the strong positive pressure wavefront crossing the same region between the time $ t = 1.5\ \si{\micro\second} $ and $ t = 2.1\ \si{\micro\second} $, which can hinder the growth of the newly created cavitation bubbles.

\subsection{Pressure sign inversion measurement}
The combined results from the numerical simulations and the experimental nucleation maps support the manifestation of Gouy phase shift during shock wave focusing. However, both represent an indirect proof of the phenomenon. Therefore, a direct measurement of the sign inversion of the pressure wave inside the droplet is sought.
The small size of the droplets used in this study (typically less than 1~\si{\milli\meter} in radius) and the high pressures expected at the focal position prevent a direct measurement of the pressure field with a commercial hydrophone. The use of the background-oriented schlieren (BOS) method overcomes this limitation by providing non-invasive measurements of the density field, which is directly related to the pressure field by means of the equation of state given in Eq.~(\ref{eq:PFH_EoS}), at the droplet center plane. 

\begin{figure}
    \centering
    \includegraphics[width=\linewidth]{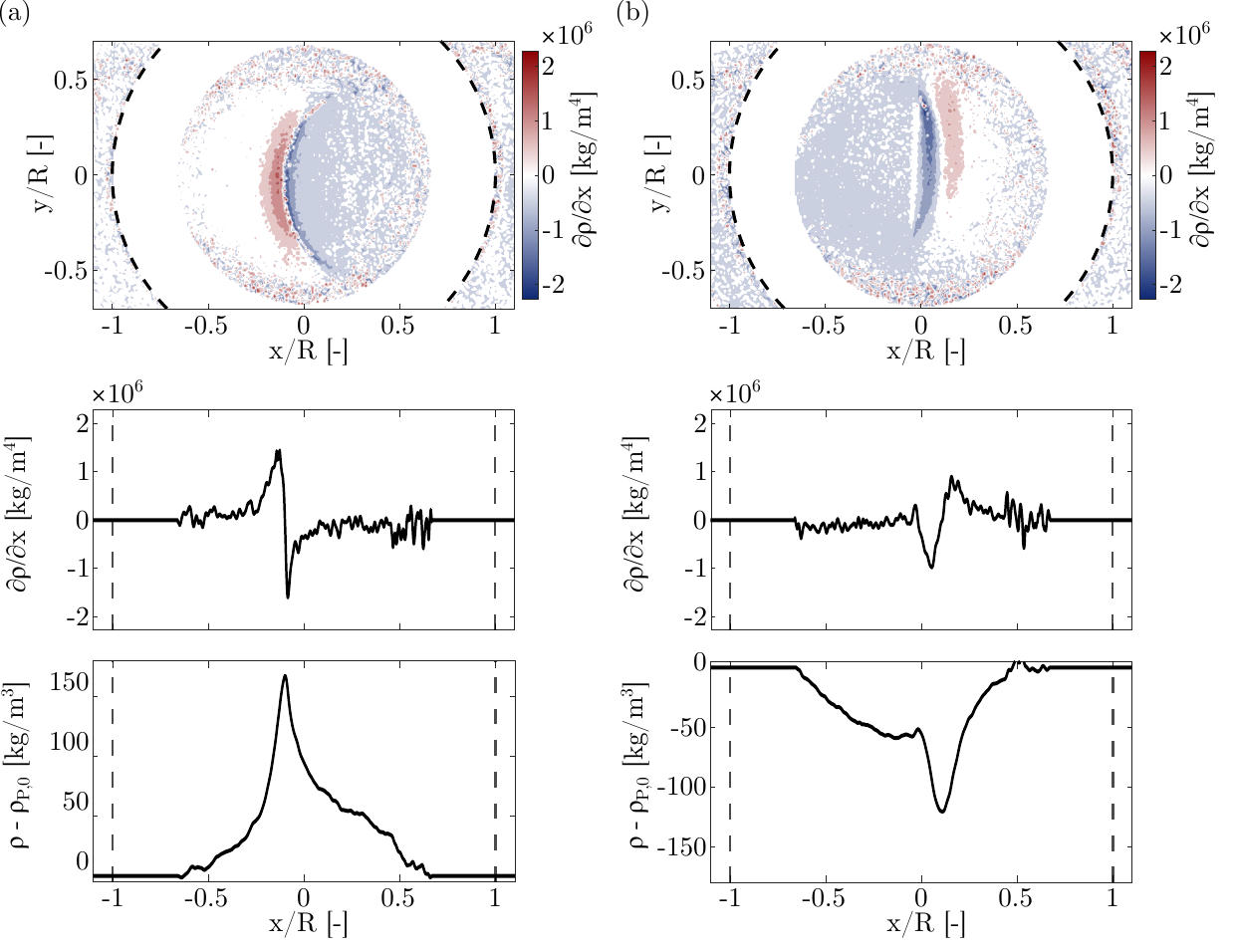}
    \caption{Density gradient field on the $xy$-plane passing through the droplet center (top) and x-axis (center) and the density variation on the x-axis (bottom) for two different instants of the shock wave propagation. The reference value for the density is $ \rho_\mathrm{P, 0} = 1680.1\ \si{\kilogram\per\meter\cubed} $. All the graphs on column (a) are obtained from a delay $ \Delta t $ = 6.1~\si{\micro \second} between the shock wave generation and the camera exposure. For column (b), a delay $ \Delta t $ = 10.4 \si{\micro \second} is employed.}
    \label{fig:BOS_integratedData}
\end{figure}

The density gradient field $ \partial \rho_{\mathrm{P}}/\partial\mathrm{x} $ measured with BOS is shown in Fig.~\ref{fig:BOS_integratedData}(a)-(b), both on the $xy$-plane passing through the droplet center (top) and on the correspondent $x$-axis (center), for two different time instants. The corresponding density along the centerline is illustrated at the bottom. In Fig.~\ref{fig:BOS_integratedData}(a), the shock wave is traveling from the left (proximal) to the right (distal) side of the droplet and has not reached the focal point yet. In Fig.~\ref{fig:BOS_integratedData}(b), the shock wave has been focused and reflected at the distal side of the droplet, and is crossing its center for the second time.

The wave displays a positive density (pressure) pulse before reaching the distal focal point [Fig.~\ref{fig:BOS_integratedData}(a)].
After the focusing and reflection at the distal interface, the wave presents a mostly negative density (pressure) pulse. As discussed above, the change in sign cannot be caused by the reflection due to the impedance mismatch. Therefore, the obtained results further support the hypothesis that the generation of negative pressure is due to the Gouy phase shift occurring during the focusing process.

\subsection{Bubble formation under the hypothesis of heterogeneous cavitation}    

Starting from the pressure field data extracted from the numerical simulations, it is possible to investigate the driving mechanism behind bubble formation inside the prefluorocarbon droplet.
Heterogeneous cavitation is typically assumed to be the leading cause of acoustic cavitation in water due to the presence of nanometric cavitation nuclei \cite{mancia_acoustic_2021}. Instead, in previous studies on perfluorocarbon droplets vaporization the appearance of vapor bubbles in the droplet bulk has typically been regarded as homogeneous nucleation \cite{lajoinie_high-frequency_2021, qin_Predicting_2021}, an assumption partly justified by the fact that droplets are usually kept in a superheated state. The droplets employed in this work are generated at ambient temperature ($ T_\mathrm{w} \simeq 298 \si{\kelvin} $), far from the PFH boiling temperature of $ T_{\mathrm{PFH}} = 329 \si{\kelvin} $ (measured at ambient pressure). Therefore, the occurrence of heterogeneous cavitation due to pre-existent nanobubbles is possible and should also be considered in a comprehensive discussion on cavitation incipience. Following the approach of \citet{mancia_acoustic_2021}, a stabilized, air-filled nanobubble population inside the PFH droplet with an initial radius $ R_0 = 3\ \si{\nano\meter} $ is hypothesized and their radial dynamics is modeled using a modified version of the Keller-Miksis equation [see Eq.~(\ref{eq:modified_Keller_Miksis})]. 

\begin{figure}[ht]
    \centering
    \includegraphics[width=\linewidth]{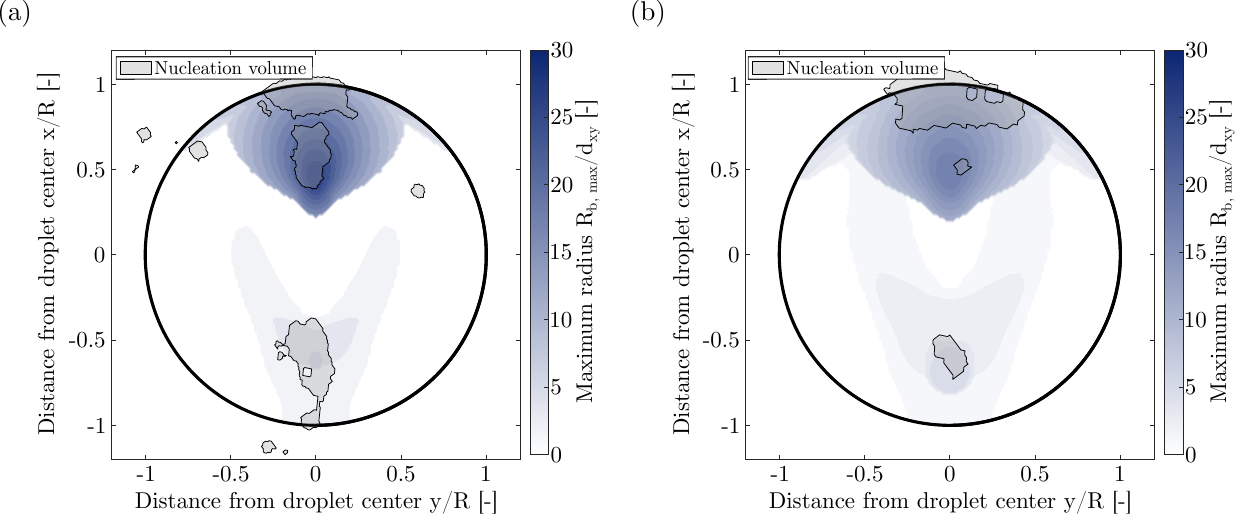}
    \caption{Maximum radius reached by a nanobubble with initial radius $ R_0 = 3~\si{\nano\meter} $ driven by the simulated pressure field in each location $ (x/R, y/R) $ in the center plane of a PFH droplet. The maximum radius $ R_{\mathrm{b, max}} $ is normalized by the pixel resolution $ d_\mathrm{xy} \simeq 8\ \si{\micro\meter} $. Results are shown for a droplet of radius (a) $ R = 825~\si{\micro\meter} $ and (b) $ R = 470~\si{\micro\meter} $. 
    The light gray areas overlapped on the graph represent the overall nucleation region detected throughout the recordings.}
    \label{fig:heterogeneousTheory}
\end{figure}

For each position $ (x,y) $ inside the droplet, the pressure field is extracted from the numerical simulation and used as the driving pressure for the bubble dynamics equation. The maximum radius $ R_{\mathrm{b, max}} $ reached by the expanded bubble nuclei is used to create the contour map displayed in Fig.~\ref{fig:heterogeneousTheory}, where its value is normalized to the pixel resolution $ d_\mathrm{xy} \simeq 8\ \si{\micro\meter}$. This means that in a location with a predicted maximum radius greater than one, cavitation is expected to be detected. 


A decent agreement is found when comparing directly these results with the experimental nucleation region, although the numerical data greatly overestimate the area of interest, especially at the distal side of the droplet. In addition, the predicted area of visible bubble activity would be even larger if the computed maximum bubble radii were considered. Small nucleation regions are detected outside the colored areas in Fig.~\ref{fig:heterogeneousTheory}(a), but the instant of their appearance in the radiograph suggests that their location is outside of the droplet. Interestingly, the negative pressure generated at the focus and transmitted outside the droplet is intense enough to initiate cavitation in the surrounding water.

In agreement with the simulations results showed in Fig.~\ref{fig:0006Simulation}, this model predicts cavitation activity close to the focus for the $ R = 470\ \si{\micro\meter} $ case (Fig.~\ref{fig:heterogeneousTheory}(b)). However, no cavitation bubbles are detected in the radiographs in the instant after wave focusing ($ t = 0 $ in Fig.~\ref{fig:0006Simulation}) at $ x/R = 0.5 $. A nucleation region near the focal point $ x/R = 0.5 $ is shown in the nucleation map obtained throughout the recording (Fig.~\ref{fig:heterogeneousTheory}(b)), but appears too late in the radiographs to be caused by the pressure field generated by the first passage of the shock wave. Probably, cavitation is generated outside the droplet or by one of the successive reflections, which are not taken into account in the present calculation.

\subsection{Bubble formation under the hypothesis of homogeneous nucleation}
\begin{figure}[ht]
    \centering
    \includegraphics[width=\linewidth]{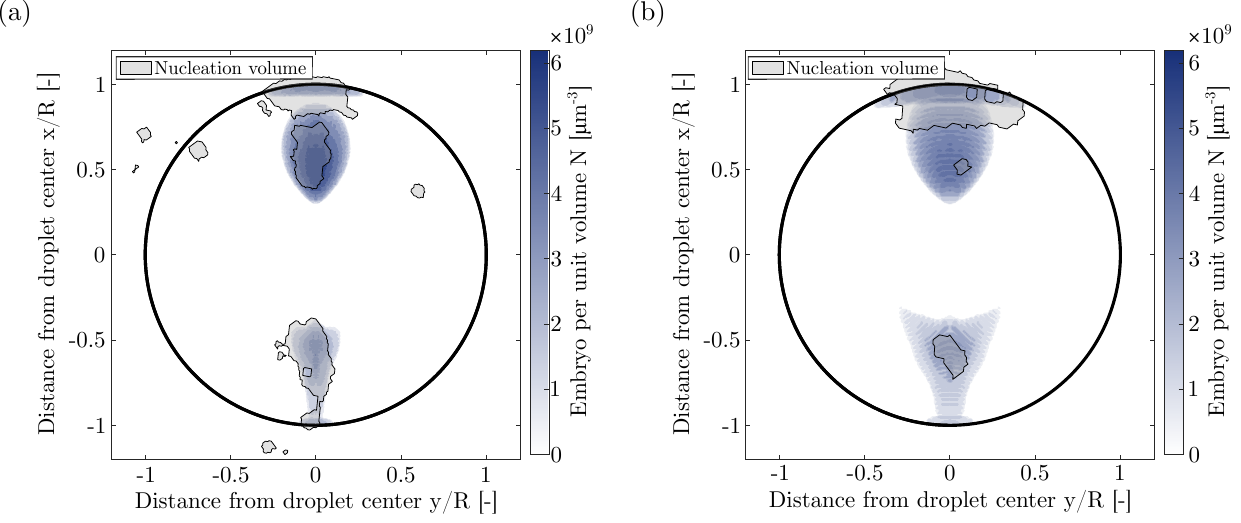}
    \caption{Number of critical nuclei generated during the interaction between a shock wave and a droplet with radius (a) $ R = 825\ \si{\micro\meter} $ and (b) $ R = 470\ \si{\micro\meter} $. The overall nucleation region (light grey areas) is overlapped with the graph to show the capability of the modified-CNT to predict the occurrence of cavitation.}
    \label{fig:homogeneousTheory}
\end{figure}

If the hypothesis of homogeneous nucleation is made, the pressure field obtained from the numerical simulations can be used to calculate the rate $ J(x, y, t) $ in every position inside the droplet for each time instant [see Eq.~\eqref{eq:HomogenousNucleation_J}]. In general, the lower the pressure in the liquid (provided that it remains below the saturation pressure), the higher the value of $ J(x, y, t) $, with a consequently higher probability of nucleation. Due to the highly unsteady nature of the shock wave propagation, we believe that not only the instantaneous nucleation rate is important to determine the likeliness of cavitation inception, but also the amount of time that the liquid remains in that state. To account for this effect, a parameter $ N(x, y) $ is defined as follows:
\begin{equation}
    N(x, y) = \int_{0}^{t_{\mathrm{f}}} J(x, y,t^\prime)dt^\prime,
    \label{eq:integrated_rate}
\end{equation}
where $ t_{\mathrm{f}} $ is the final simulation time. $ N(x, y) $ simply represents the amount of nuclei per unit volume generated over a certain time span. Therefore, higher cavitation activity is expected for larger values of this parameter. Fig.~\ref{fig:homogeneousTheory} shows the comparison between $ N(x, y) $ and the experimental nucleation region for both the droplet sizes.  

The best agreement between the nucleation map and the estimated value of the nuclei generated per unit volume is for the case of a $ R = 825\ \si{\micro\meter} $ droplet, presented in Fig.~\ref{fig:homogeneousTheory}(a). The three main areas of nucleation, corresponding to region near the focal point at $ x/R = 0.5 $, the interface of the distal side ($ x/R = 1$), and the elongated cavitation region on the proximal side respectively, are well predicted by the parameter $ N(x, y) $. In contrast to the heterogeneous cavitation model, the parameter $ N(x, y) $ accurately represents the two distinct cavitation regions located at the droplet distal side. The larger size of the detected region at $ x/R = 1$ is probably due to the fact that the experimental cavitation map takes into account the maximum expansion of the cavitation bubble, while $ N(x, y) $ only represents the initial nucleation spot.

For the case of the $ R = 470\ \si{\micro\meter} $ droplet shown in Fig.~\ref{fig:homogeneousTheory}(b), the homogeneous nucleation model correctly shows the nucleation region around $ x/R = 1 $ and describes well the increase of its (relative) size compared to the previous case. The cavitation region on the proximal side is also in good agreement with experiments, with the detected bubbles being at the location of the local maximum of $ N(x,y) $. However, experimental results do not show nucleation around the focal point $ x/R = 0.5 $. 
As discussed in Section~\ref{sec:numericalResults}, the discrepancy might be caused by uncertainties in the shock wave pressure profile.

\section{Conclusions}\label{sec:Conclusions}
This work demonstrates localized, negative pressure generation through the focusing of a rarefaction-free shock wave using a perfluorohexane droplet as acoustic lens. Both numerical simulations and projected-background oriented schlieren measurements are performed to study the shock wave propagation within the droplet. Since the impedance mismatch between the two materials does not introduce any sign inversion when the wave front in the droplet is reflected at the perfluorocarbon-water interface, the emergence of tension is completely attributed to the Gouy phase shift phenomenon. 
High-speed x-ray phase-contrast recordings show bubble nucleation inside the droplet, confirming that the generation of negative pressure by phase shift can be exploited for controlled initiation of cavitation. Homogeneous nucleation is identified as the main physical mechanism behind bubble nucleation, ensuring that the results are repeatable and do not rely on dissolved gasses or other impurities inside the droplet. 

The choice of a perfluorocarbon droplet as an acoustic lens directly links the geometry studied here to ADV applications. The occurrence of Gouy phase shift in the droplet could justify the use of shock wave generation devices as a safer alternative to ultrasonic transducers to vaporize the therapeutic agents, although more accurate studies on the scalability of the phase shift effect are needed to extend the presented results the sub-micrometer scale.

Beyond the specific configuration presented in this study, the occurrence of Gouy phase shift can be extended for any type of pressure wave in a liquid, provided that it is focused on a single point.
The presented findings can therefore inspire novel acoustic driving strategies for cavitation-based therapeutic applications that exploit the generation of focused positive pressure pulses. The absence, or the reduction, of externally supplied rarefaction pressures can reduce off-target cavitation activity, making the techniques safer and more precise, while the reduction of the number of cycles with respect to periodic ultrasonic driving decreases energy deposition away from the focal region, reducing heating and contributing to the prevention of unwanted side-effects.

\begin{acknowledgments}
The authors acknowledge the financial support from the Swiss National Science Foundation (Grant No.\ 200567), ETH Zürich, the Tokyo University of Agriculture and Technology, and the access to beamtime at beamline ID19 of the European Synchrotron Radiation Facility in the frame of the Shock BAG MI-1397. 
S.I.\ and Y.T.\ acknowledge the support by the Japan Society for the Promotion of Science KAKENHI (Grant Nos.\ JP22KJ1239, JP24H00289); the Japan Science and Technology Agency PRESTO (Grant No.\ JPMJPR21O5); Heiwa Nakajima Foundation.
The authors also thank Dr. Gazendra Shakya for the assistance in the preliminary experiments and data processing, Dr.~Arnaud Sollier for the support during the ESRF experimental campaign and Dr.~Kevin Schmidmayer for the help in setting up the ECOGEN simulations.
\end{acknowledgments}

\bibliography{references.bib}

\newpage
\appendix 
\section{Triggering scheme}
\label{app:trigger}
\begin{figure}[h]
    \centering
    \includegraphics[width = \linewidth]{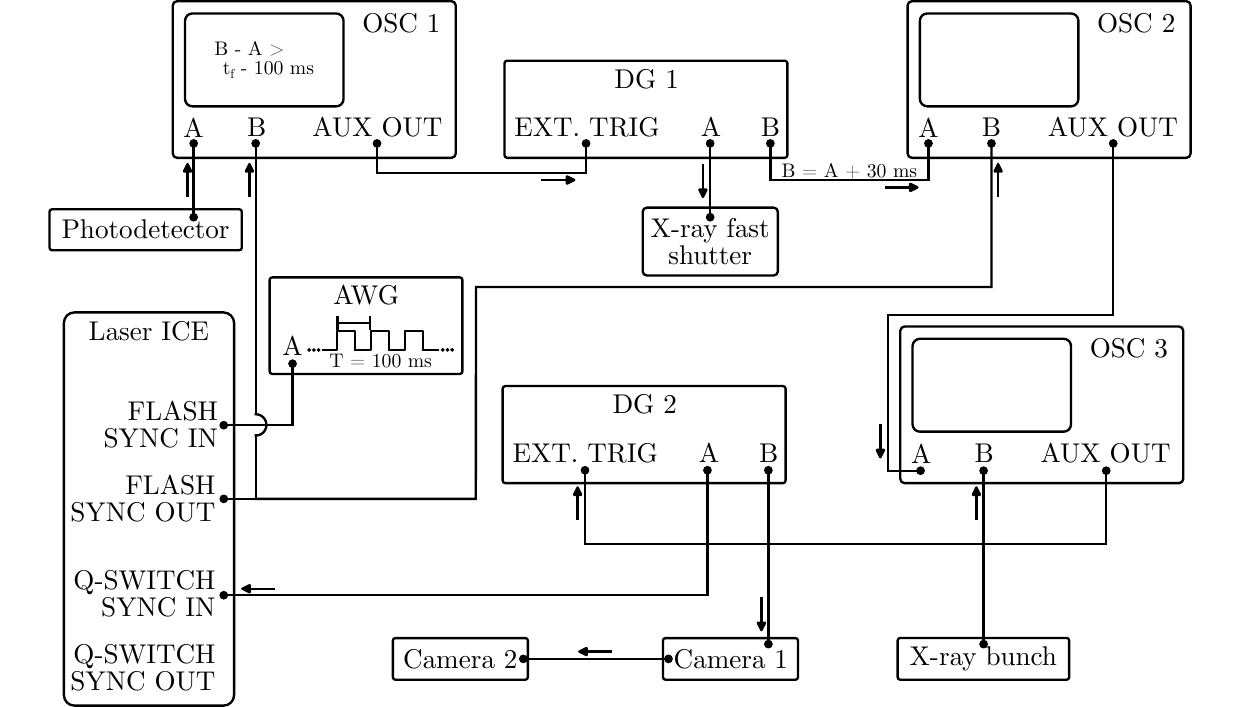}
    \caption{Block diagram of the triggering circuit used to acquire the high-speed video recordings at the ESRF. Channel A and B of the oscilloscopes (OSC A and B) are used as inputs for qualified triggers. In OSC 1 an additional condition on channel B has been added, i.e., the signal has to be detected with a delay $ \Delta t_\mathrm{O1} = t_\mathrm{f} - 100~ \si{\milli \second} $ with respect to the event in channel A. Here $ t_\mathrm{f} $ is the droplet falling time, measured experimentally. Channel A and B of delay generators (DG A and B) are set to send a TTL pulse to one of the other instruments. If not specified otherwise, the delay of the DG output with respect to the EXTERNAL TRIG input has to be considered equal to zero.}
    \label{fig:experimentalSetupTrigger}
    \end{figure}

    The experimental setup shown in Fig.~\ref{fig:experimentalSetupXRay} requires a complex triggering system, meant to ensure that the cameras fully record the interaction between the PFH droplet and the shock wave. Moreover, the cameras' exposure time needs to be synchronized with the x-ray bunch to provide the best possible illumination.
    A schematic of the triggering sequence is presented in Fig.~\ref{fig:experimentalSetupTrigger}. In this configuration, the laser's Integrated Cooling and Electronics (ICE) unit requires two input signals for operation, one to synchronize the laser flashlamp (periodic square pulse), and one to trigger the Q-Switch and shoot the laser. An Arbitrary Wave Generator (AWG) provides a continuous 10~\si{\hertz} TTL pulse train to the Flashlamp Sync In (FLASH SYNC IN) input of the laser ICE, in order to keep the laser ready to shoot at the moment a TTL pulse is read at the Q-SWITCH SYNC IN port.

    The full sequence can be described as follows:

    \begin{itemize}
        \item \textbf{Droplet detection}: a laser diode aimed at a photodetector measures the passage of the droplet across the light path via a voltage drop detected by channel A of an oscilloscope (OSC 1).
        
        \item \textbf{Laser synchronization}: the Flashlamp Sync Out (FLASH SYNC OUT), synchronized with the flashlamp trigger output, is connected to channel B of OSC 1. 
        Using a qualified trigger function, a TTL pulse is sent from the AUX OUT channel of OSC 1 when the voltage drop on channel A and the FLASH SYNC OUT pulse on channel B are detected, only if the event in channel B occurs at least $ \Delta t_\mathrm{O1} = t_\mathrm{f} - 100~\si{\milli \second}$ after the one in channel A.
        Here, $ t_\mathrm{f} $ is the droplet falling time from the capillary to the camera field of view (measured experimentally) and 100~\si{\milli \second} is the period T of the FLASH SYNC IN signal. 
        Since the x-ray shutter opening time is relatively long ($ \approx 20~\si{\milli \second} $), shooting the laser right after detecting the signal in channel B is not possible. The delay ensures that the droplet will be in the cameras' field of view at the next available pulse.
        
        \item \textbf{X-ray shutter opening}: The signal from the channel AUX OUT of OSC 1 is used to trigger a delay generator (DG 1) which immediately sends a TTL pulse to the x-ray shutter and a second pulse to channel A of a second oscilloscope (OSC 2), with a delay $ \Delta t_\mathrm{DG1} \approx 30~\si{\milli \second} $ to accommodate for the shutter opening time. 
        A TTL pulse is sent from the AUX OUT channel of OSC 2 after both the signal in channel A and B are detected consecutively, the latter being connected with the FLASH SYNC OUT port of the laser ICE. The signal from AUX OUT is now synchronized with the first laser flashlamp pulse available after the shutter opening. 
        
        \item \textbf{X-ray bunch synchronization}: 
        Channel A and B of a third oscilloscope (OSC~3) are connected to the AUX OUT port and the x-ray bunch trigger signal respectively. 
        A TTL pulse is sent from the AUX OUT ports of OSC~3 when both the signals on channel A and subsequently channel B are detected and is connected to the external trigger port (EXT. TRIG) of a delay generator (DG 2).
        
        \item \textbf{Cameras and laser triggering}: Once triggered, DG 2 produces two separate pulses, connected to the Q-SWITCH SYNC OUT port of the laser ICE and to the CAMERA 1 trigger input. A delay $ \Delta t = 172\ \si{\micro \second} + \Delta t_\mathrm{QS} $ is applied to both the outputs, where the $ 172\ \si{\micro \second} $ is needed to take into account internal delays in the ICE and $ \Delta t_\mathrm{QS} $ is the Q-Switch delay value controlling the laser energy, fixed to $ 272\ \si{\micro \second} $. 
        CAMERA 2 is controlled by CAMERA 1, ensuring synchronization between the two recordings. 

    \end{itemize}

    \section{Supplementary videos}
    \label{sec:suppMat}
    The experimental recordings and the animations obtained from the numerical simulation are available for download and their content is described below: 

    \begin{itemize}
        \item The files \textbf{825umDroplet\_XRay.avi} and \textbf{470umDroplet\_XRay.avi} contain the x-ray phase-contrast videos, obtained as described in Section~\ref{sec:xrayMethods}, for a droplet of radius $ R = 825\ \si{\micro\meter} $ and $ R = 470\ \si{\micro\meter} $ respectively. A selection of the frames is reported in Fig.~\ref{fig:experimentalXRaySnapshots} for both sizes. 

        \item The files \textbf{825umDroplet\_Shadowgraph.avi} and \textbf{470umDroplet\_Shadowgraph.avi} contain the shadowgraph videos, recorded following the method described in Section~\ref{sec:xrayMethods}, for a droplet of radius $ R = 825\ \si{\micro\meter} $ and $ R = 470\ \si{\micro\meter} $ respectively. For each droplet size, a frame in which nucleation is visible is shown in Fig.~\ref{fig:experimentalXRaySnapshots}. The presence of the shock wave in the proximal side of the droplet is visible in \textbf{825umDroplet\_Shadowgraph.avi} at time t = -2.3 \si{\micro\second} as a thin line appearing in the droplet center as well as a change in the lighting. The same feature can be seen in \textbf{470umDroplet\_Shadowgraph.avi} at time t = -1.77 \si{\micro\second}.

        \item The files \textbf{825umDroplet\_Simulation.avi} and \textbf{470umDroplet\_Simulation.avi} contain the animation of the simulated pressure field for a droplet of radius $ R = 825\ \si{\micro\meter} $ and $ R = 470\ \si{\micro\meter} $ respectively, calculated using the method detailed in Section~\ref{sec:simulationSetup}.
    \end{itemize}
    In every video, the overlayed time stamps show the delay between the current frame and the instant in which nucleation is first detected. The arrows labeled \enquote{SW} indicate the shock wave propagation direction.
\end{document}